\documentclass[manuscript]{aastex62}
\usepackage[encapsulated]{CJK}
\usepackage{subfigure}
\usepackage{mathrsfs}
\usepackage{amsmath}
\usepackage{epstopdf}
\usepackage{longtable}
\usepackage{booktabs}
\usepackage{tablefootnote}
\usepackage{threeparttable}
\usepackage{multirow}
\usepackage{hyperref} %\usepackage{threeparttable}
\def\etal {et al.~}

\newbox\grsign \setbox\grsign=\hbox{$>$} \newdimen\grdimen \grdimen=\ht\grsign
\newbox\laxbox \newbox\gaxbox
\setbox\gaxbox=\hbox{\raise.5ex\hbox{$>$}\llap
	{\lower.5ex\hbox{$\sim$}}}\ht1=\grdimen\dp1=0pt
\setbox\laxbox=\hbox{\raise.5ex\hbox{$<$}\llap
	{\lower.5ex\hbox{$\sim$}}}\ht2=\grdimen\dp2=0pt
\shorttitle{Light Deflection by Jupiter}
\shortauthors{Li \etal}

\begin{document}
\begin{CJK*}{UTF8}{gbsn}

	\title{Light Deflection under the Gravitational Field of Jupiter --- Testing General Relativity}
		
	\correspondingauthor{Yingjie Li}
	\email{liyj@pmo.ac.cn, xuye@pmo.ac.cn}
		
	\author{Yingjie Li}\affiliation{Purple Mountain Observatory, Chinese Academy of Sciences, Nanjing 210023, China}
	\author{Ye Xu}\affiliation{Purple Mountain Observatory, Chinese Academy of Sciences, Nanjing 210023, China}
	\affiliation{University of Science and Technology of China, Hefei, Anhui 230026, China}
		
	\author{JingJing Li}
	\affiliation{Purple Mountain Observatory, Chinese Academy of Sciences, Nanjing 210023, China}
		
	\author{Yuanwei Wu}
	\affiliation{National Time Service Center, Key Laboratory of Precise Positioning and Timing Technology, Chinese Academy of Sciences, Xi'an 710600, People's Republic of China}
		
	\author{Shaibo Bian}
	\affiliation{Purple Mountain Observatory, Chinese Academy of Sciences, Nanjing 210023, China}
	\affiliation{University of Science and Technology of China, Hefei, Anhui 230026, China}	
	
	\author{ZeHao Lin}
	\affiliation{Purple Mountain Observatory, Chinese Academy of Sciences, Nanjing 210023, China}
	\affiliation{University of Science and Technology of China, Hefei, Anhui 230026, China}	
	
    \author{WenJin Yang}
    \affiliation{Max-Planck-Institut f\"{u}r Radioastronomie, Auf dem H\"{u}gel 69, 53121, Bonn, Germany}
	
	\author{Chaojie Hao}
    \affiliation{Purple Mountain Observatory, Chinese Academy of Sciences, Nanjing 210023, China}	
	\affiliation{University of Science and Technology of China, Hefei, Anhui 230026, China}    
    
    \author{DeJian Liu}
    \affiliation{Purple Mountain Observatory, Chinese Academy of Sciences, Nanjing 210023, China} 
	\affiliation{University of Science and Technology of China, Hefei, Anhui 230026, China}    	

\begin{abstract}
We measured the relative positions between two pairs of compact extragalactic sources (CESs), J1925-2219 \& J1923-2104 (C1--C2) and J1925-2219 \& J1928-2035 (C1--C3) on 2020 October 23--25 and 2021 February 5 (totaling four epochs), respectively, using the Very Long Baseline Array (VLBA) at 15 GHz. Accounting for the deflection angle dominated by Jupiter, as well as the contributions from the Sun, planets other than Earth, the Moon and Ganymede (the most massive of the solar system's moons), our theoretical calculations predict that the dynamical ranges of the relative positions across four epochs in R.A. of the C1--C2 pair and C1--C3 pair are 841.2 and 1127.9 $\mu$as, respectively. The formal accuracy in R.A. is about 20 $\mu$as, but the error in Decl. is poor. The measured standard deviations of the relative positions across the four epochs are 51.0 and 29.7 $\mu$as in R.A. for C1--C2 and C1--C3, respectively. These values indicate that the accuracy of the post-Newtonian relativistic parameter, $\gamma$, is $\sim 0.061$ for C1--C2 and $\sim 0.026$ for C1--C3. Combining the two CES pairs, the measured value of $\gamma$ is $0.984 \pm 0.037$, which is comparable to the latest published results for Jupiter as a gravitational lens reported by Fomalont \& Kopeikin, i.e., $1.01 \pm 0.03$.
\end{abstract}

%\tableofcontents

\section{Introduction} 

The first observation of light deflection by the Sun \citep{Dyson+1920}, which tested the theory of general relativity, is praised as the beginning of the ``modern era'' \citep{Johnson1983}. Testing general
relativity has been a perennial hot issue since then, and with the accumulation of astrometric data and technological developments, increasingly better accuracy is being achieved. The deflection of light and the theories behind it, including relativistic gravity and gravitational lensing theory, have become important tools for both astronomy and cosmology, including studying galactic structure, probing dark matter and dark energy, hunting for fluctuations in the cosmic microwave background radiation, and finding exoplanets and dwarf stars, etc. \citep{Will2015}.

In the coming decades another 3--4 orders of magnitude of improvement in testing relativistic gravity is expected to be achieved, taking advantage of the rapid development of modern technology \citep{Ni2017}. Indeed, the next-generation radio observatories, e.g., the Square
Kilometre Array \citep[SKA,][]{Braun+2015} and the next-generation Very Large
Array \citep[ngVLA,][]{Murphy+2018}, and their pathfinders  are working towards ultraprecise ($\sim$ 1 $\mu$as or higher) and ultrasensitive \citep[e.g., a few tenths $\mu$Jy/beam @ 9.2 GHz for an 8 hr exposure with SKA,][]{Bonaldi+2021} astrometry \citep{Rioja-Dodson2020}. To date, astrometric accuracy is dozens of $\mu$as or better. For example, the typical parallax accuracy of the Bar and Spiral Structure Legacy (BeSSeL) Survey is $\sim$ 20 $\mu$as \citep{Reid+2019}; the current accuracy of the 3rd realization of the International Celestial Reference Frame (ICRF3) is $\sim$ 30 $\mu$as \citep{Charlot2020}; and the typical positional accuracy of bright stars with $G < 15$ mag is $\sim$ 20 $\mu$as in Gaia's Early Data Release 3 \citep{Gaia-Collaboration+2021}. To date, the record parallax precision is $\pm$ 3 $\mu$as \citep{Zhang+2013}. It is essential to test different metric theories of gravity \citep[e.g.,][]{Damour-Nordtvedt1993,Fomalont+2009}, and this in turn is vital to future ultraprecise and ultrasensitive observatories, because the difference between the predicted deflection angles of the different theories may be comparable to or even smaller than the expected astrometric errors. Likewise, it is equally essential to test higher-order parameterized post-Newtonian (PPN, hereafter) formalism \citep[e.g.,][]{Crosta-Mignard2006, Kopeikin-Makarov2007}. In short, measuring deflection angles and testing gravitational theory or its high-order approximations are two indispensable and urgent tasks.

As expressed in PPN formalisms, the dimensionless parameter $\gamma$ is used to characterize the contribution of space curvature to gravitational deflection. After more than a hundred years of hard work, the highest accuracy of $\gamma$ reached $\sim 2\times10^{-5}$ via data obtained with the Cassini spacecraft \citep{Shapiro1964,Bertotti+2003}. Very long baseline interferometry (VLBI) facilities have also made great and successive contributions in determining $\gamma$ \citep[e.g.,][]{Robertson-Carter1984, Lebach+1995, Lambert+2009}. The key research achievements are listed below. \citet{Shapiro2004} measured $\gamma - 1 = (-1.7 \pm 4.5) \times 10^{-4}$ using nearly two million VLBI observations obtained at 87 sites toward 541 radio sources, and \citet{Fomalont+2009} obtained $\gamma - 1 = (-2 \pm 3) \times 10^{-4}$ with Very Long Baseline Array (VLBA) observations at higher frequency (i.e., 43 GHz) to reduce the light bending by the electron content in the solar corona. The latest accuracy of $\gamma$ obtained using the VLBI technique is $\sim 9\times10^{-5}$ \citep{Titov+2018}. In the above studies, the Sun acts as a gravitational lens \citep[for more of the history of measuring $\gamma$, see][]{Will2015}. However, the solar corona contains large and rapid plasma-density fluctuations that limit the measured accuracy of the gravitational deflection angle. The effect of plasma is much weaker for other planets in the solar system. In this work, Jupiter is selected as a gravitational lens.  

So far, of all the planets, only Jupiter has been used to measure $\gamma$. By measuring the light deflection events in 2002 when Jupiter passed within 3.7$'$ of the quasar J0842+1835, \citet{Fomalont-Kopeikin2008} measured the PPN parameter $\gamma$ of $1.01 \pm 0.03$. The ratio between the retarded deflection and that predicted by general relativity \citep[see the details in][]{Kopeikin2001, Kopeikin-Fomalont2002} was measured as $0.98 \pm 0.19$ \citep[ see][]{Fomalont-Kopeikin2003}. This ratio can be used to evaluate the speed of propagation of gravity \citep{Kopeikin2001, Kopeikin-Fomalont2002}.

In such a background, it is a good choice to take full advantage of the weak influence of plasma in the atmospheres of the solar system's planets and natural satellites. The maximum deflection angles predicted by theory exceed $\sim$ 16 mas for Jupiter and $\sim$ 5 mas for Saturn \cite[e.g.,][]{Crosta-Mignard2006}. From the decade-long monitoring of light-bending events, the accuracy of $\gamma$ determined using deflected light from compact extragalactic sources (CESs) by planets or natural satellites is expected to be of the order of $10^{-5}$ or better. These observations could be combined with observations of light-bending events by the Sun in order to test and develop second-order PPN formalisms, to measure the speed of propagation of gravity waves, to investigate the properties of the (noninertial) motion and multi-pole moment of the gravitational field of a lens like Jupiter \citep[e.g.,][]{Kopeikin-Gwinn2000, Kopeikin+2011, Kopeikin-Makarov2021}, to probe effects in multi-plane lensing \citep[e.g.,][]{Subramanian-Chitre1984, Erdl-Schneider1993, Ramesh+2021}, to test different metric theories of gravity and to model their gravitational effects imposed on light from distant celestial bodies. Also, this would be preparation for ultraprecise and ultrasensitive astrometry to be conducted by next-generation observatories (including SKA and ngVLA, etc.). In turn, ultraprecise and ultrasensitive observations would be beneficial in terms of improving astrometric accuracy, and thus the accuracy of the parameters of PPN (e.g., $\gamma$). Such cross-fertilization could contribute to both the theoretical aspects of gravity and the greater precise of astrometry via improved methods and techniques.

The remainder of this paper is organized as follows. In Section \ref{sec:data}, we describe the data used in this work. In Section \ref{sec:results}, we present the observational results and compare them with the theoretical predictions. An estimation of $\gamma$ is given in this section. In Section \ref{sec:uncertaincy}, we discuss the influence of plasma and the positional errors of the gravitational lens. Finally, we summarize our conclusions in Section \ref{sec:summary}.

\section{Data}\label{sec:data}

\subsection{Observational Strategy}\label{sec:data_sched}

\begin{figure}[!htb]
	\centering
%	\subfigbottomskip=-0.2cm
	\subfigcapskip=-0.2cm
	\subfigure[Deflection: J1925-2219]{\includegraphics[width=0.49\textwidth]{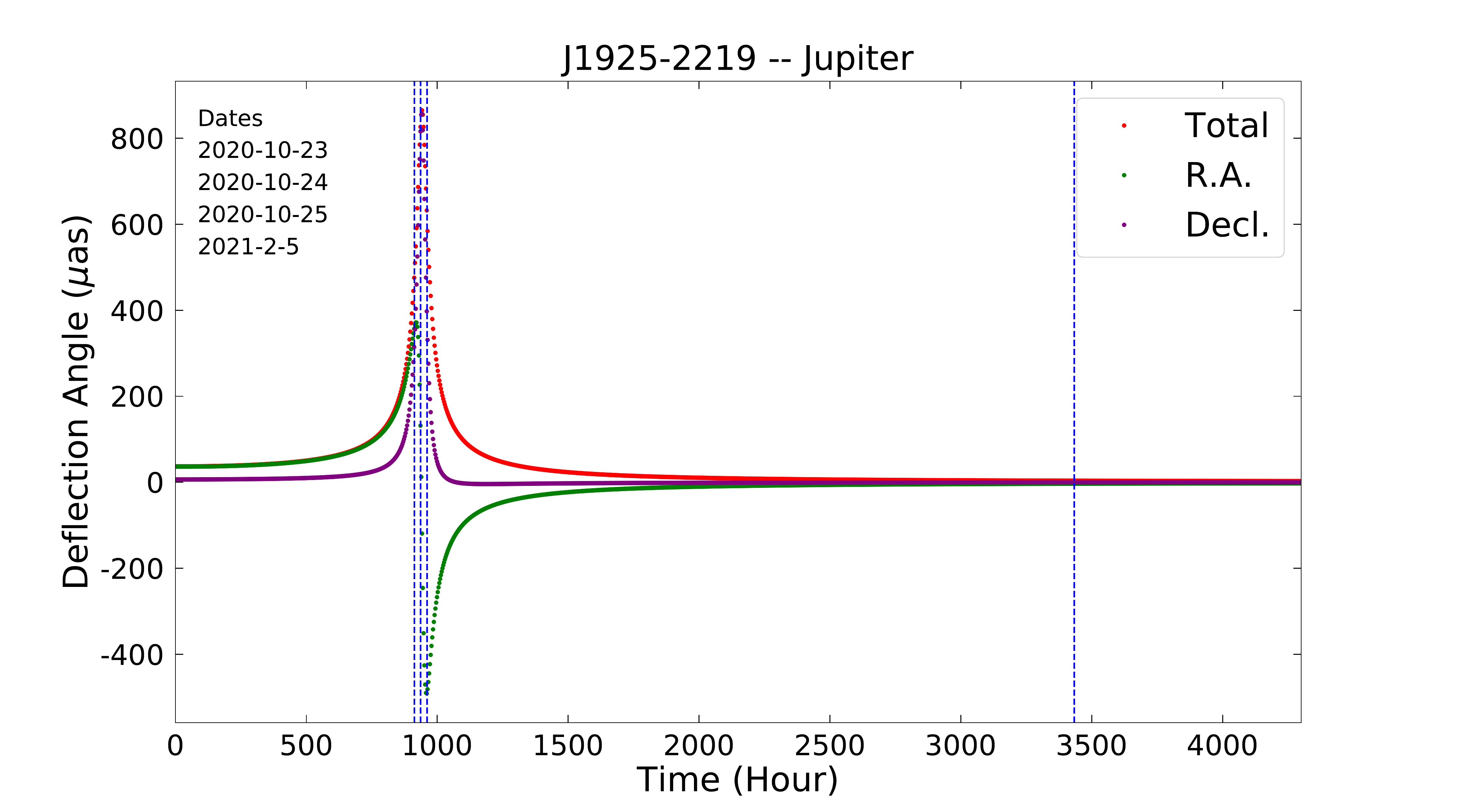}}
	\subfigure[impact parameter: J1925-2219]{\includegraphics[width=0.49\textwidth]{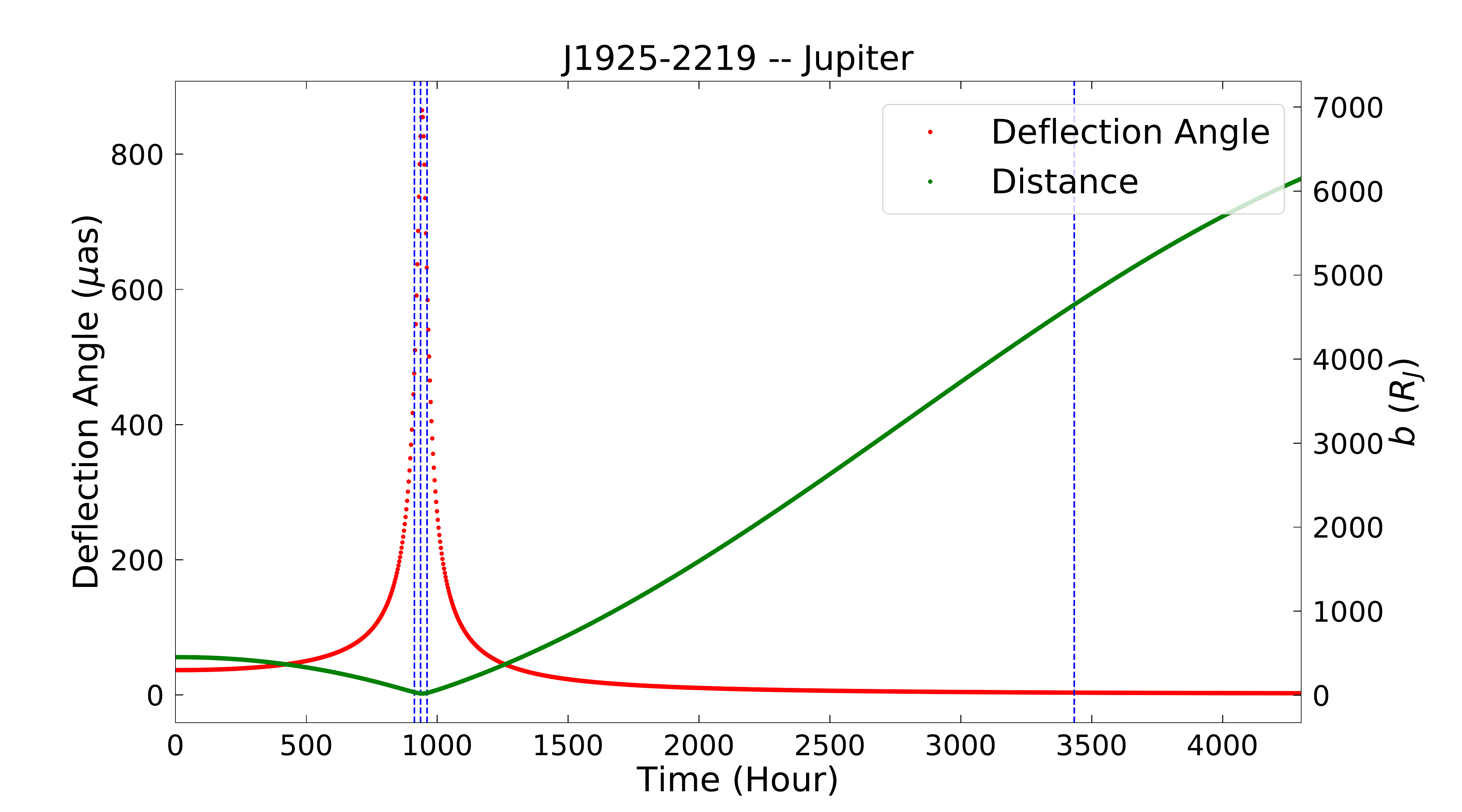}}	
	\subfigure[Deflection: J1923-2104]{\includegraphics[width=0.49\textwidth]{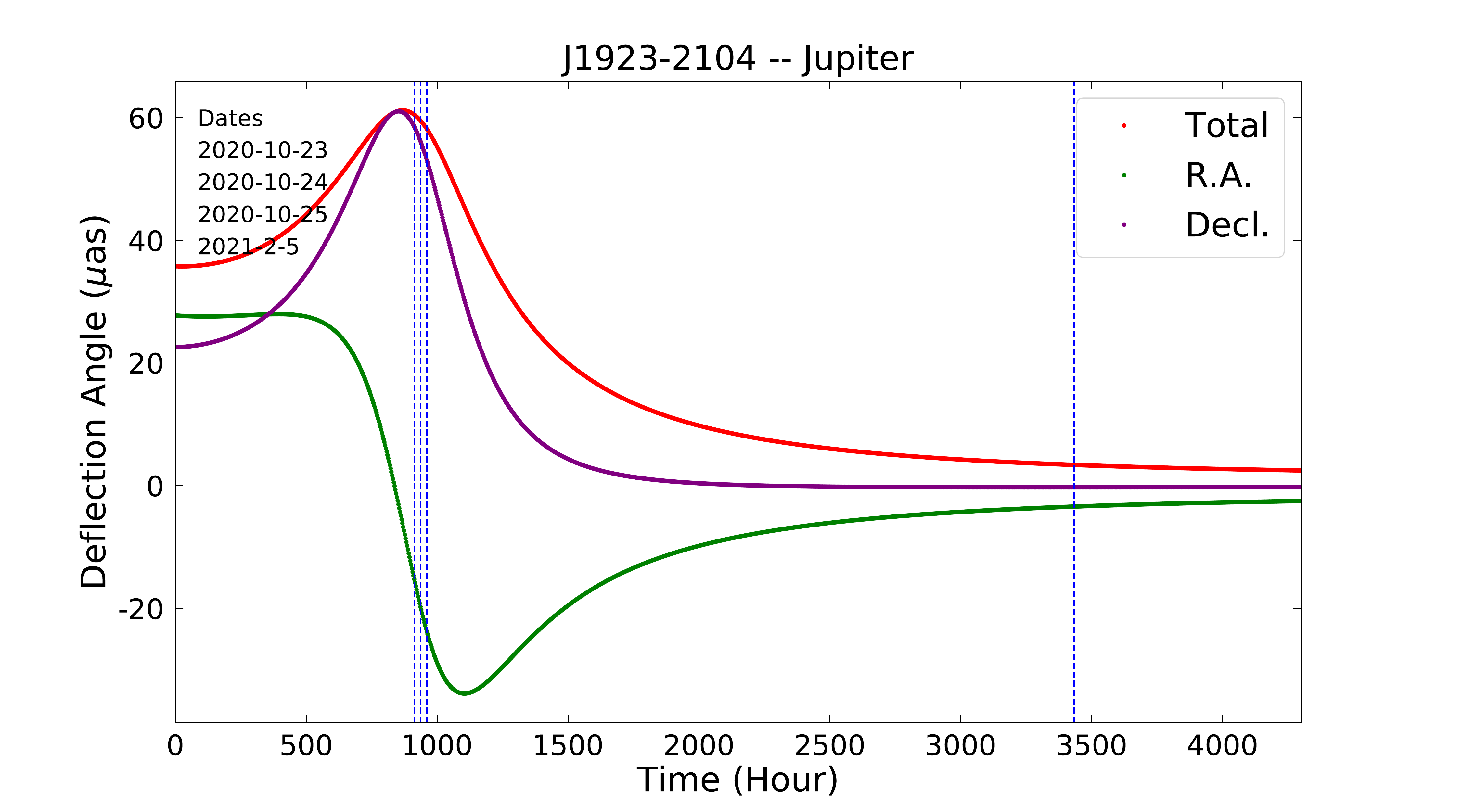}}
	\subfigure[impact parameter: J1923-2104]{\includegraphics[width=0.49\textwidth]{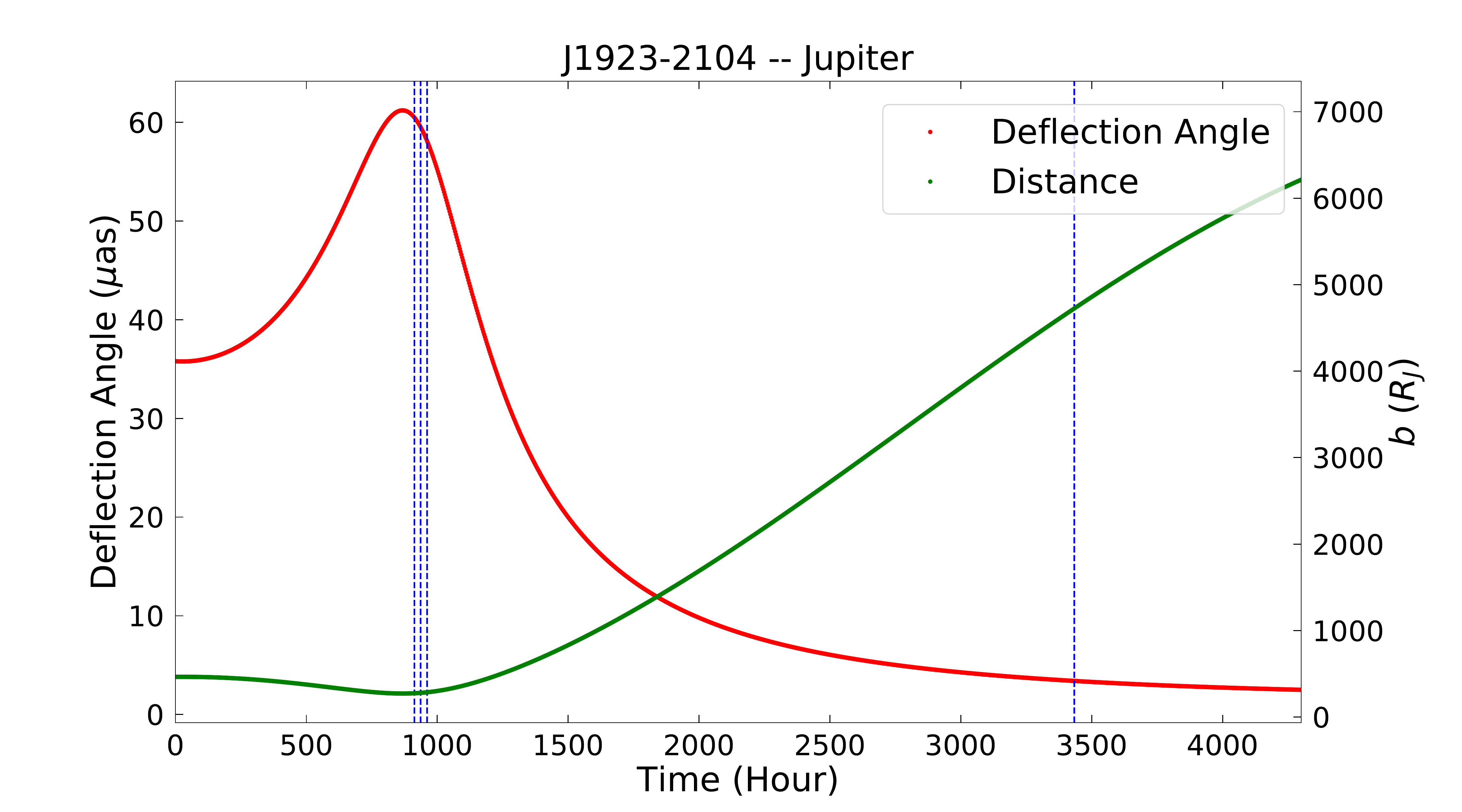}}
	\subfigure[Deflection: J1928-2035]{\includegraphics[width=0.49\textwidth]{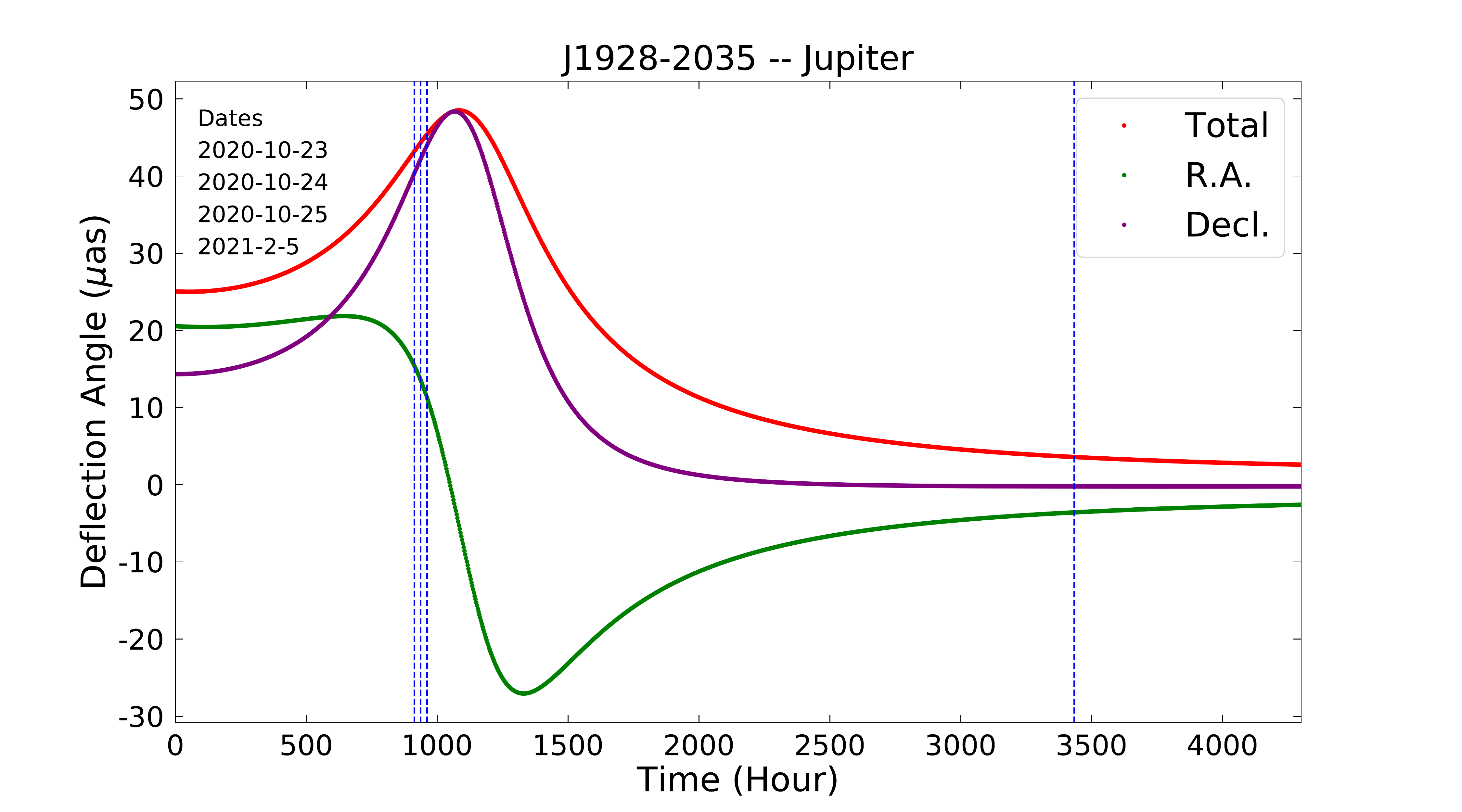}}
	\subfigure[impact parameter: J1928-2035]{\includegraphics[width=0.49\textwidth]{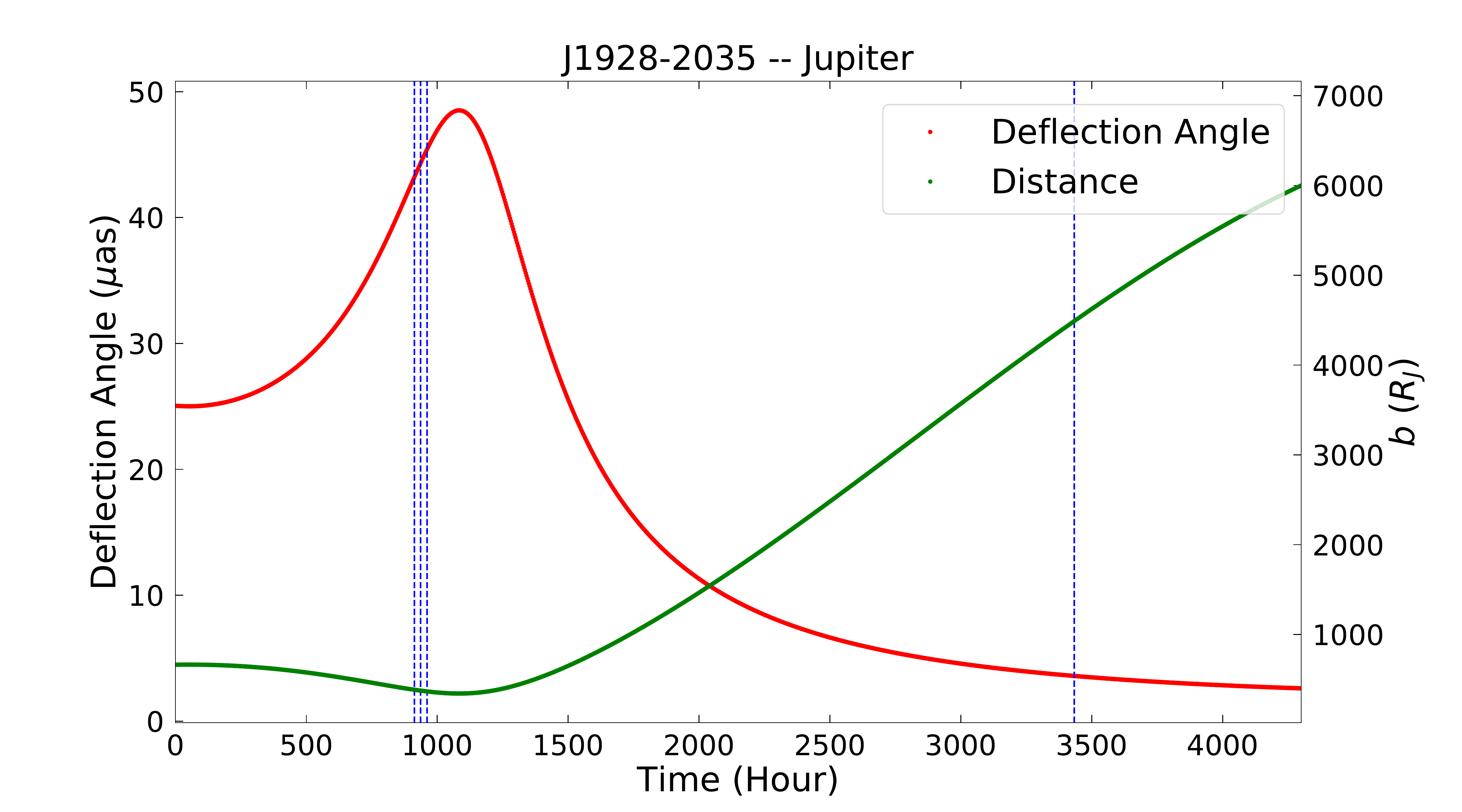}}	
	\caption{Left: the predicted light deflection angle of light of J1925-2219 (top), J1923-2104 (middle) and J1928-2035 (bottom) caused by Jupiter. The red curve denotes the total deflection angle, and the purple and green curves present the deflection angle in R.A. and Decl., respectively. Right: the predicted total deflection angle (red curve) and the corresponding impact parameter $b$ (green curve); see Section \ref{sec:delfection-all-solar-system}. The zero-point represents 00:00:00 (UT1) on 2020 September 15. The blue vertical lines represent the four epochs: 2020 October 23, 24, and 25 and 2021 February 5, GST. }
	\label{fig-design}
\end{figure}

To take full advantage of the weak influence of plasma in Jupiter's atmosphere relative to the Sun, the observational strategy revolves around Jupiter as a gravitational lens. The selected CES is J1925-2219 from ICRF3 \citep{Charlot2020}. Figure \ref{fig-design} presents the predicted deflection angle (see Section \ref{sec:delfection-all-solar-system}) of J1925-2219 and its two nearby CESs (i.e., J1923-2104 and J1928-2035, also from ICFR3) by Jupiter on 2020 October 23, 24, and 25, and 2021 February 5, GST. An observing day is considered as an epoch, where each are numbered 1 to 4 sequentially. The relative positions between the line of sight of the three CESs and Jupiter are also plotted. The first three days are selected around peak deflection angles (see panel (a) in Figure \ref{fig-design}). The component of deflection in R.A.(J2000) almost reaches a maximum and minimum during epochs 1 and 3, respectively, while that in Decl.(J2000) reaches a maximum during epoch 2. We have also added epoch 4 on 2021 February 5. In this epoch, the calculated deflection angle caused by Jupiter is $<$ 5 $\mu$as. The effect of plasma on Jupiter is less than 0.3 $\mu$as in all four epochs.

\subsection{Observations and Data Reduction}\label{sec:data_reduction}

We chose CESs J1923-2104 and J1928-2035 as background sources from ICRF3 \citep{Charlot2020}. Table \ref{table:CESs} lists the positions of these two CESs and J1925-2219. Using the National Radio Astronomy Observatory's (NRAO's)\footnote{NRAO is a facility of the National Science Foundation operated under cooperative agreement by Associated Universities, Inc.} VLBA, we observed them at 15 GHz from 2020 October 23 to 2021 February 5 for a total of four epochs (see above), under program BL277. Each epoch contains a 3.5 hr track.

\begin{deluxetable}{cccccccc}
	\centering
    \tabletypesize{\footnotesize}
	\setlength\tabcolsep{3pt}
	\tablecolumns{8}
	\tablewidth{15cm}
	\tablecaption{Detailed Information about the Three CESs
	\label{table:CESs}}
	%\begin{tabular}
	\tablehead{
		\colhead{Name} & \colhead{Index} & \colhead{R.A.(J2000)} & \colhead{Decl.(J2000)} &  \colhead{R.A.$_{\mathrm{sep}}$} & \colhead{Decl.$_{\mathrm{sep}}$} & \colhead{Beam Size} & \colhead{$F_{\mathrm{peak}}$ } \\
		\colhead{} & \colhead{} & \colhead{$\mathrm{(^h\;\;\;^m\;\;\;^s})$} & \colhead{$(^\circ\;\;\;'\;\;\;'')$} & \colhead{(deg)} & \colhead{(deg)} & \colhead{(mas, mas, deg)} & \colhead{(Jy beam$^{-1}$)}
	}
	\startdata
	J1925-2219 & C1 & 19 25 39.790185 & -22 19 35.11262 & ... & ... & 6.5 $\times$ 3.0 @ 4 & 0.155 \\
	J1923-2104 & C2 & 19 23 32.189814 & -21 04 33.33312  & 0.49 & -1.25 &  6.5 $\times$ 3.0 @ 4 & 0.280 \\ 
	J1928-2035  & C3 & 19 28 09.183357 & -20 35 43.78454 & 0.53 & -1.73 & 6.5 $\times$ 3.0 @ 4 & 0.103 \\
	\enddata
	\tablewidth{15cm}
\tablecomments{The R.A.$_{\mathrm{sep}}$ is $\delta$R.A. $\cos$(Decl.), where $\delta$R.A. is the offset of the R.A. between the target C1 and the calibrators (C2 and C3) listed in the table. The beam size and $F_{\mathrm{peak}}$ are from the first epoch.}
\end{deluxetable}

Three geodetic blocks were put at the start, middle and end of each track, and phase-referenced observations were inserted between them. The observational setup and calibration procedures were similar to those used by \citet{Reid+2009-mar}, except that: only continuum sources are included in this work; the frequency is 15 GHz; and four dual-polarized intermediate-frequency (IF) bands of 64 MHz, correlated with 128 channels per IF, were employed for the observations. All of the data were processed with the DiFX\footnote{DiFX is the Swinburne University of Technology's software correlator for VLBI, which was developed as part of the Australian Major National Research Facilities Programme, and is operated under licence.} correlator (see a brief description of the principles in Section \ref{sec:delfection-all-solar-system}) at the VLBA correlation facility in Socorro, New Mexico, USA \citep{Deller+2007}. 

We used J1925-2219 as the phase-reference source. After mapping the three CESs, the task {\it{JMFIT}} in the NRAO's Astronomical Image Processing System \citep[AIPS;][]{van-Moorsel+1996} was used to fit the two-dimensional Gaussian brightness distributions, yielding their relative positions. 

\section{Results and Analyses}\label{sec:results}

\subsection{Observational Results}\label{sec-results-observational}

\begin{figure}[!htb]
	\centering
	\includegraphics[width=0.49\textwidth]{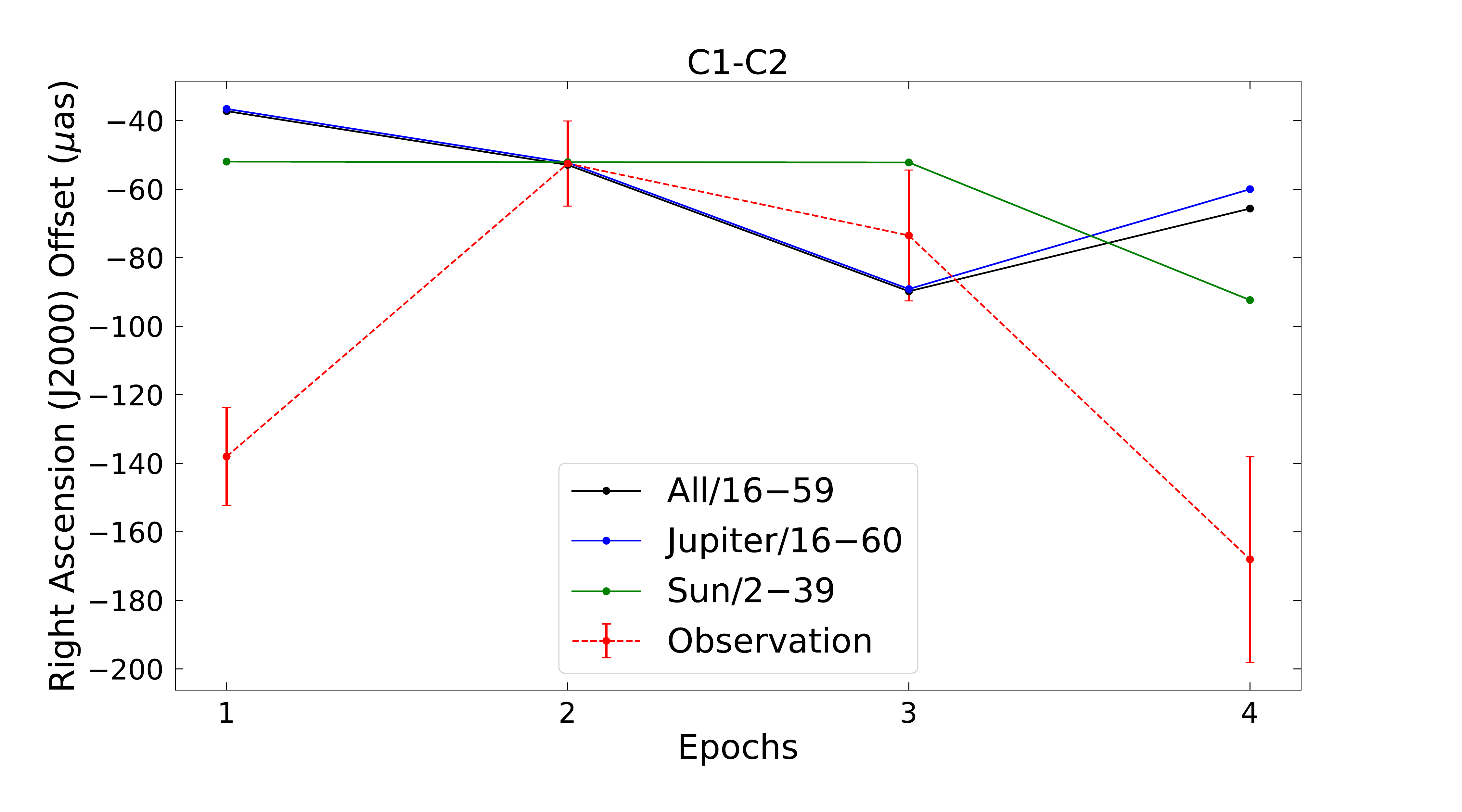}
	\includegraphics[width=0.49\textwidth]{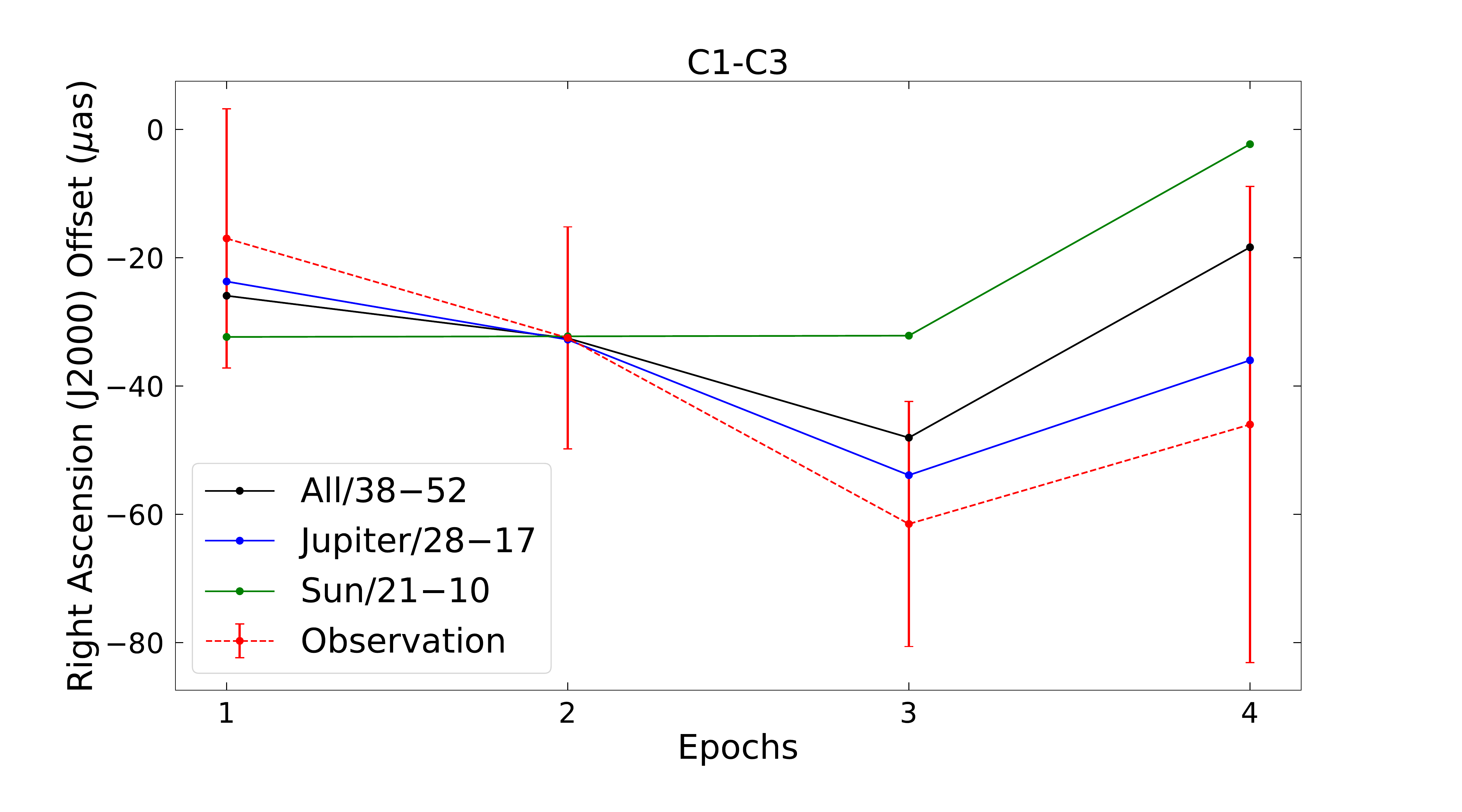}	
	\caption{Comparison of the observed relative positions between the CES pairs (left: C1--C2; right: C1--C3) with the theoretical values. To make the comparison easier, the dynamical ranges of the observational and theoretical calculations (``All'', ``Jupiter'', and ``Sun'', see Table \ref{tab:alpha-theory}) have been set to be the same as the observational results; i.e., the theoretical results were multiplied by Std/DR, where the corresponding ``DR'' values are given in Table \ref{tab:alpha-theory} (see below) and the ``Std'' values in Table \ref{tab:alpha-result}. The revised theoretical values were further adjusted to match the observational values in epoch 2. Such treatments are labeled in the legend, e.g., ``All/16$-$59'' denotes the theoretical value (see Table \ref{tab:alpha-theory}) divided by 16 (DR/Std) and minus 59~$\mu$as.}
	\label{fig-alpha-observal}
\end{figure}

The observed relative positions between the two CES pairs (i.e., J1925-2219 and J1923-2104, C1--C2, as well as J1925-2219 and J1928-2035, C1--C3) are shown in Table \ref{tab:alpha-result} and Figure \ref{fig-alpha-observal}. These relative positions are relative to the coordinates used to correlate the VLBA data (see Table \ref{table:CESs}), where the errors of these coordinates in R.A. are denoted as $C_{C1,C2}$ for C1--C2 and $C_{C1,C3}$ for C1--C3 (see the details in Section \ref{sec:result-gamma}). Because these three sources have low Decl. (i.e., as low as $\sim -22^{\circ}$; see Table \ref{table:CESs}), the components in this direction have poor relative positions, largely due to systematic errors from unmodeled atmospheric delays. Therefore, only the components in R.A. were chosen to analyze the parameter $\gamma$. The formal accuracy in R.A. is about 20 $\mu$as. The standard deviations of the changes in the relative positions throughout all the epochs are 51.0 $\mu$as for CES pair C1--C2 and 29.7 $\mu$as for C1--C3 in R.A. 

\begin{deluxetable}{cccccccccccccc}
	\centering
	\tabletypesize{\scriptsize}
	\setlength\tabcolsep{2.5pt}
	\tablecolumns{14}
	\tablewidth{15cm}
	\renewcommand{\arraystretch}{1.2}
	\tablecaption{Observed Relative Positions for Each CES Pair for All Four Epochs
		\label{tab:alpha-result}}
	%\begin{tabular}
	\tablehead{
		%		\\[0.05mm]
		\colhead{Index} & \multicolumn{6}{c}{In R.A.} & & \multicolumn{6}{c}{In Decl.} \\
		\cline{2-7}  \cline{9-14}		
		& \colhead{Epoch 1} & \colhead{Epoch 2} & \colhead{Epoch 3} & \colhead{Epoch 4} & \colhead{Mean} & \colhead{Std} & & \colhead{Epoch 1} & \colhead{Epoch 2} & \colhead{Epoch 3} & \colhead{Epoch 4} & \colhead{Mean} & \colhead{Std} \\
		& \colhead{($\mu$as)} & \colhead{($\mu$as)} & \colhead{($\mu$as)} & \colhead{($\mu$as)} & \colhead{($\mu$as)} & \colhead{($\mu$as)} & & \colhead{($\mu$as)} & \colhead{($\mu$as)} & \colhead{($\mu$as)} & \colhead{($\mu$as)} & \colhead{($\mu$as)} & \colhead{($\mu$as)}
	}
	\startdata
	C1--C2	&	-138.0$\pm$14.3	&	-52.5$\pm$12.4	&	-73.5$\pm$19.1 	&	-168.0$\pm$30.1 	& -108.0 & 51.0	&  &	-1127.0$\pm$28.6 	&	-1070.5$\pm$18.7 	&	-990.5$\pm$33.4	&	-1348.5$\pm$88.5	& -1134.1 & 142.1	\\
	C1--C3	&	-17.0$\pm$20.2 	&	-32.5$\pm$17.3 	&	-61.5$\pm$19.1 	&	-46.0$\pm$37.1 	& -39.3	&	29.7 &	& -437.0$\pm$37.5 	&	-624.0$\pm$25.5	&	-552.0$\pm$40.3 	&	-655.0$\pm$113.4 	& -567.0	 &  105.7 \\
	\enddata
	\tablewidth{15cm}
	\tablecomments{Std. denotes standard deviation.}
\end{deluxetable}

\subsection{Theoretical Results}\label{results-theoretical-intro}

\subsubsection{Overview of the Theoretical Computations}\label{sec:delfection-all-solar-system}

\begin{figure}[!htb]
	\centering
	\includegraphics[width=0.49\textwidth]{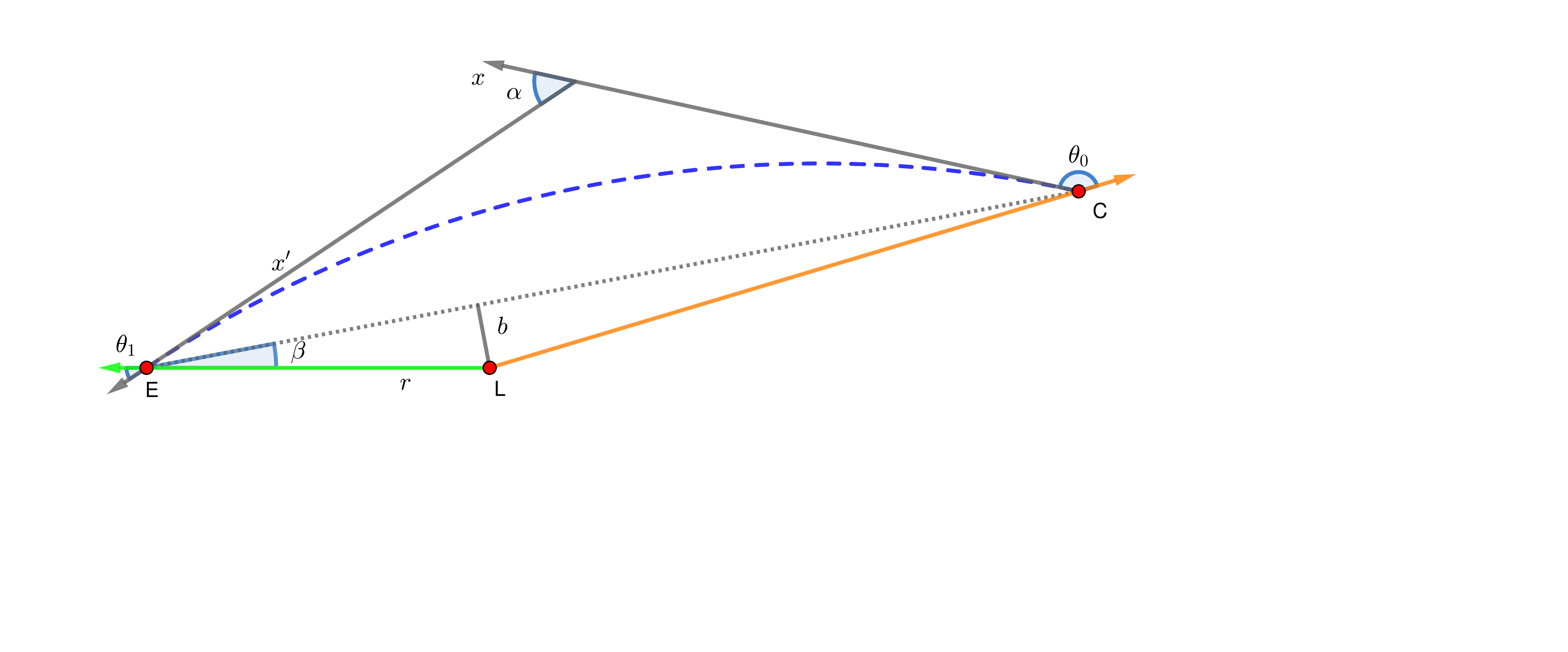}
	\includegraphics[width=0.49\textwidth]{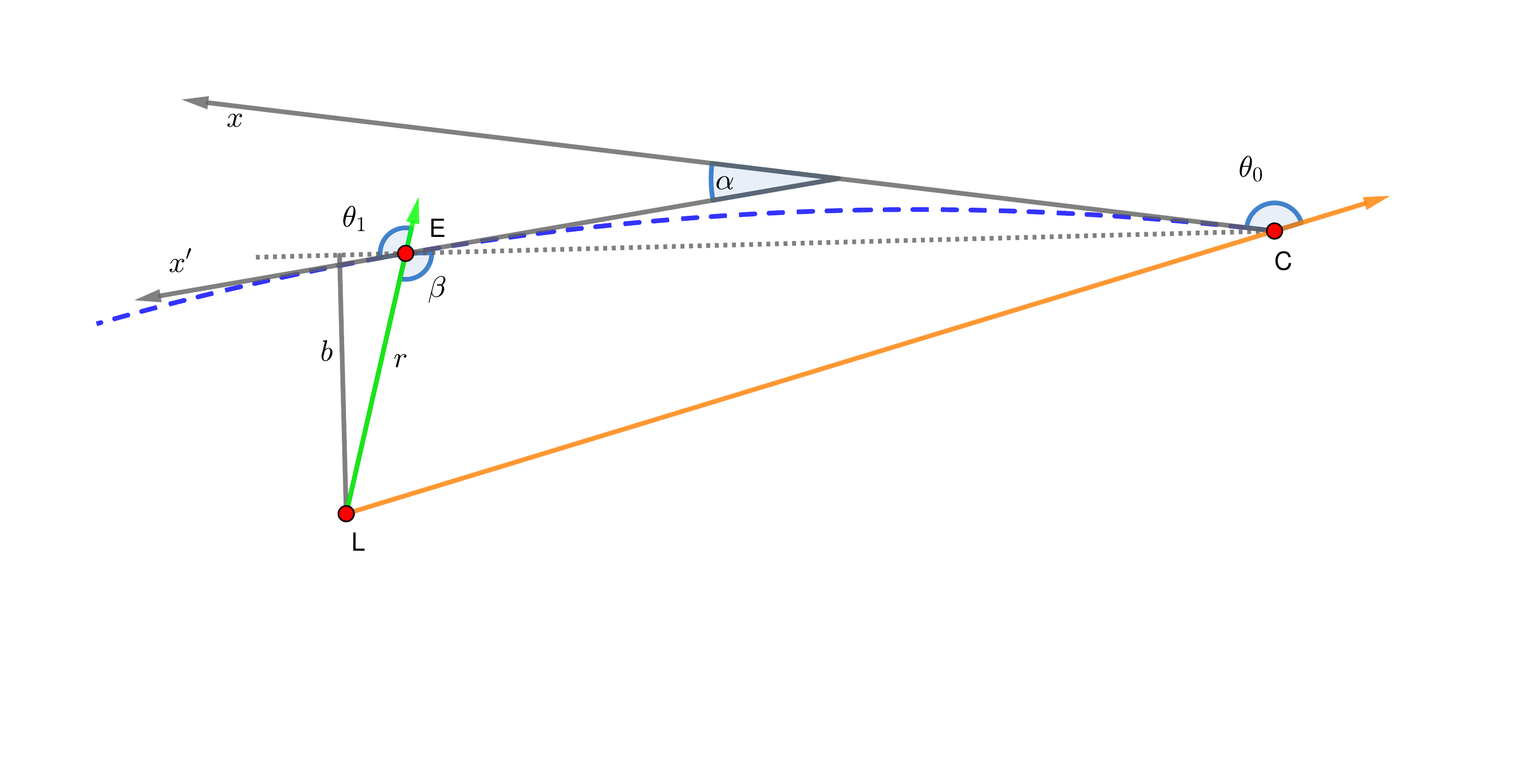}	
	\caption{Schematic diagrams of light deflection under the gravitational field of a lens using GeoGebra (\url{https://www.geogebra.org}, Copyright \copyright~International GeoGebra Institute, 2013). Left: $\beta < 90^{\circ}$; Right: $\beta > 90^{\circ}$. L, C, and E denote the gravitational lens, the CES, and Earth, respectively. The $x$ and $x'$ axes present the axis of the initial direction of light propagation and the direction after bending, respectively.}
	\label{fig-skeptical}
\end{figure}

\begin{figure}[!htb]
	\centering
	\subfigbottomskip=-0.2cm
	\subfigcapskip=-0.2cm
	\subfigure[Jupiter with Maximum Distance from Earth]{\includegraphics[width=0.49\textwidth]{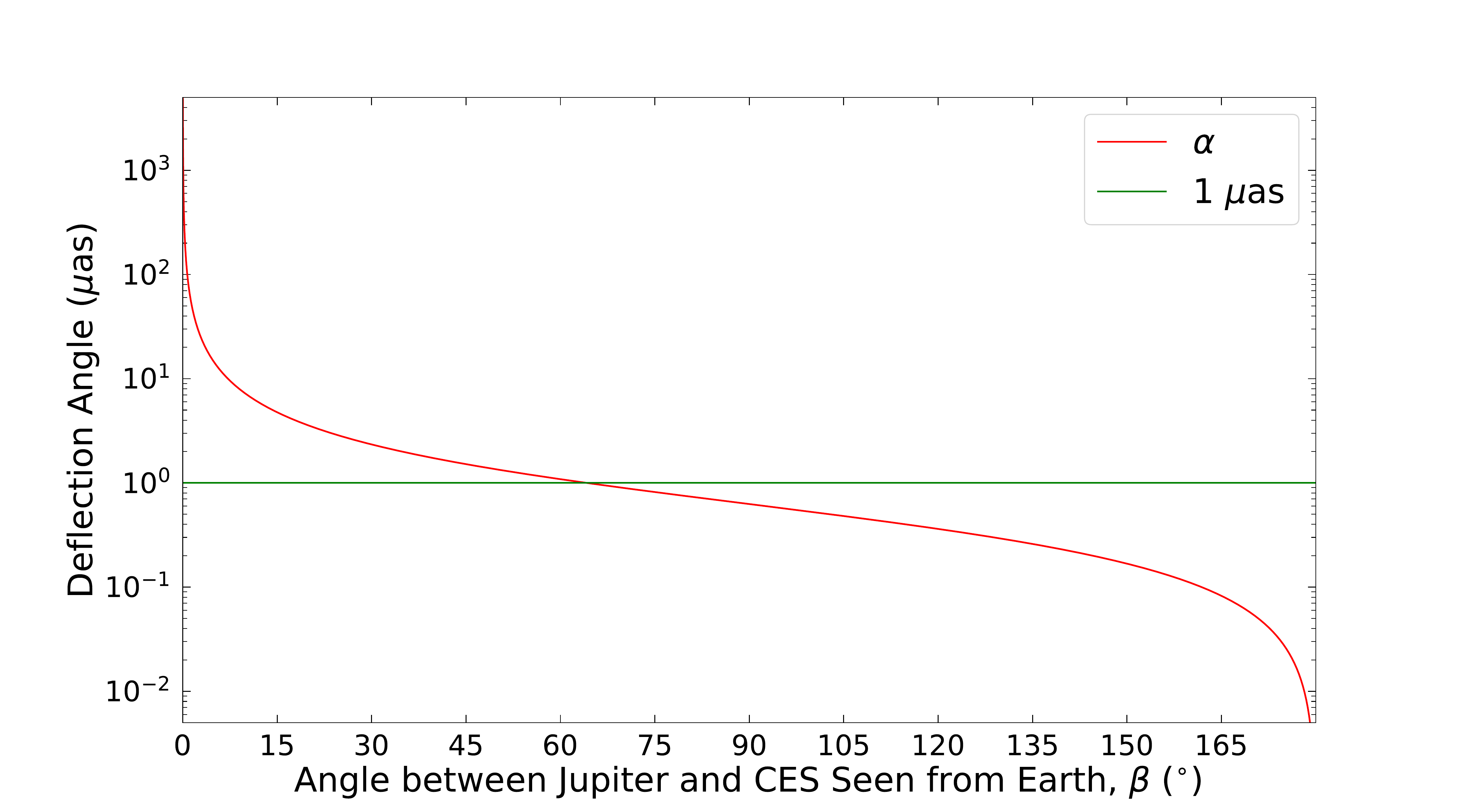}}
	\subfigure[Jupiter with Minimum Distance from Earth]{\includegraphics[width=0.49\textwidth]{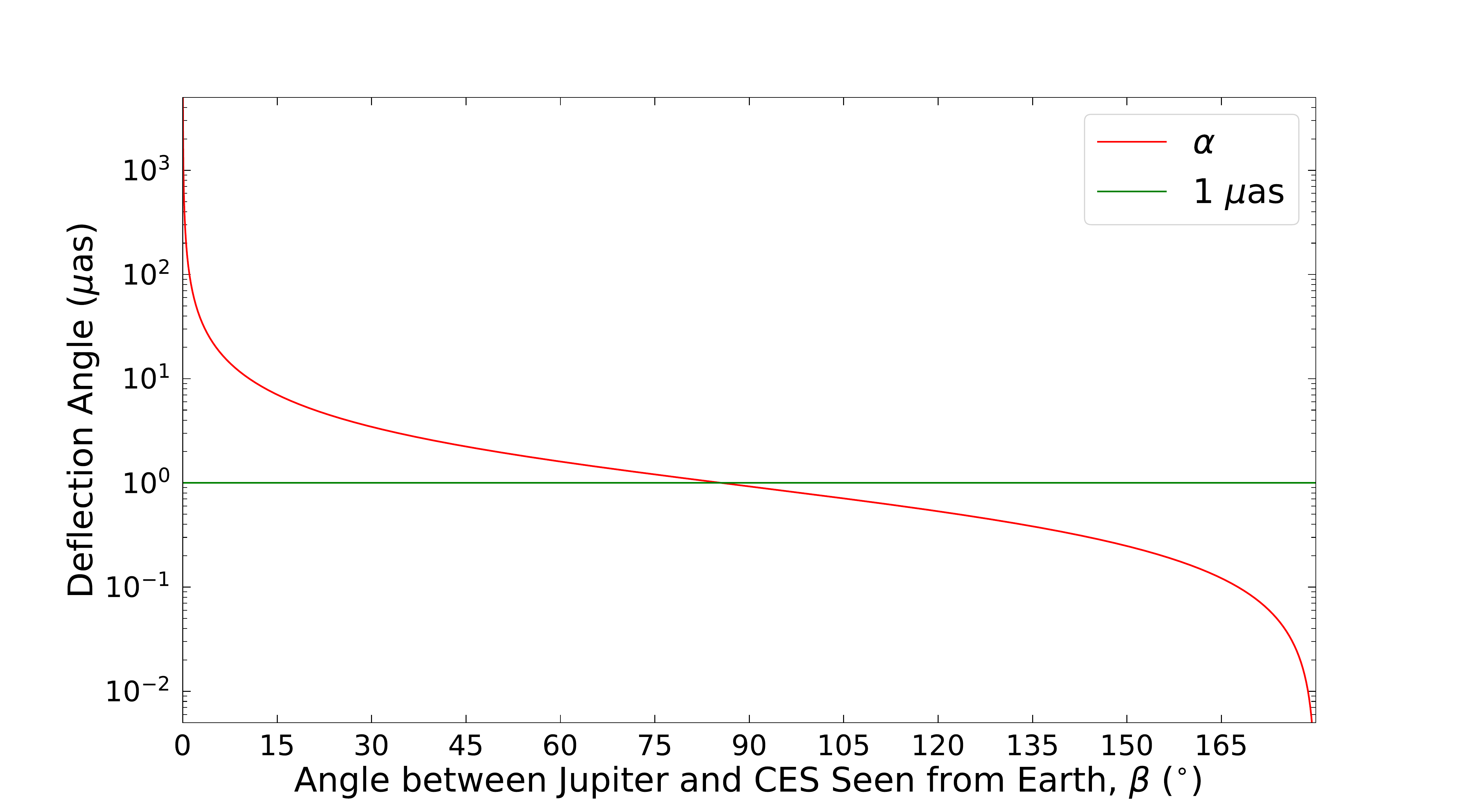}}	
	\subfigure[Saturn with Maximum Distance from Earth]{\includegraphics[width=0.49\textwidth]{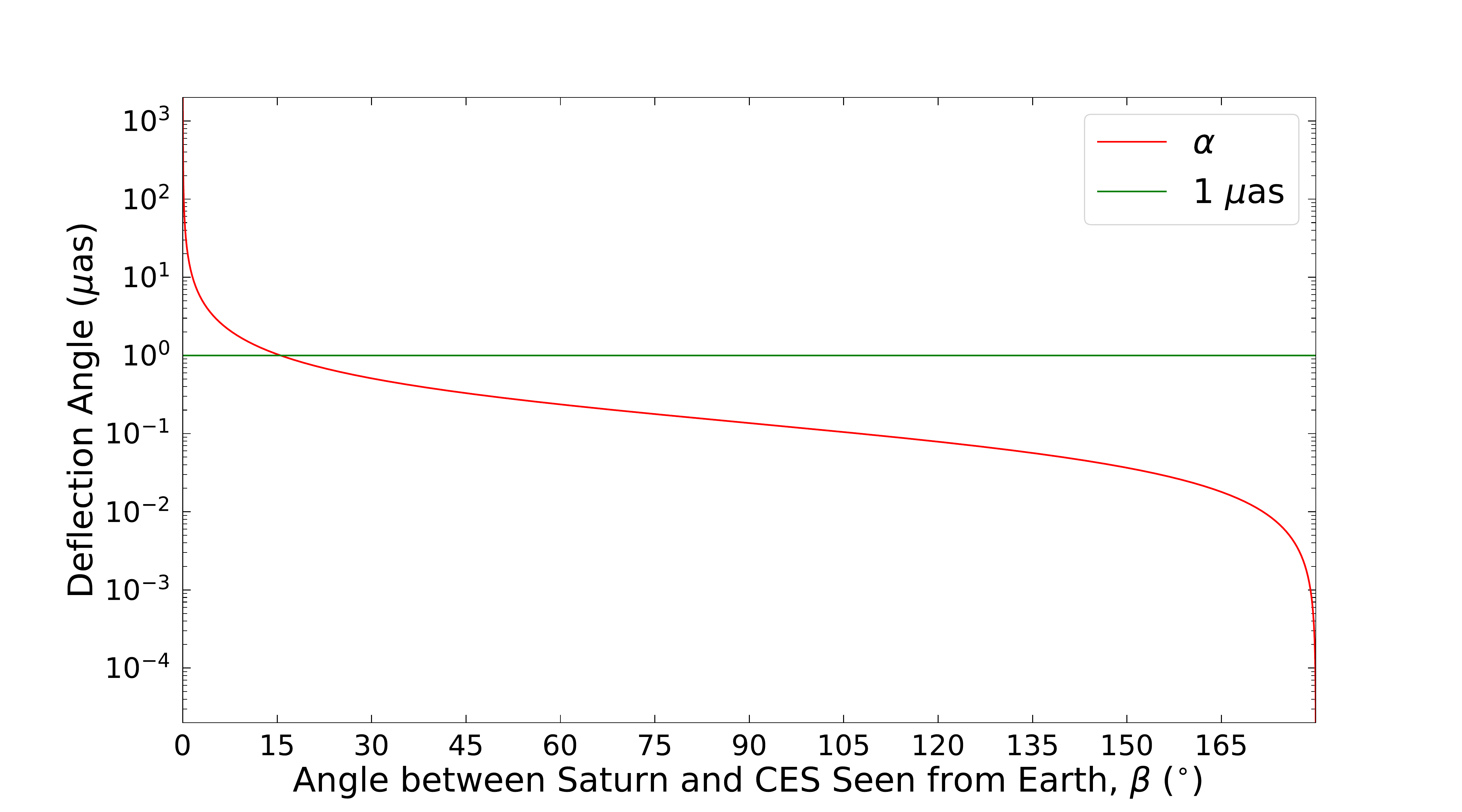}}
	\subfigure[Saturn with Minimum Distance from Earth]{\includegraphics[width=0.49\textwidth]{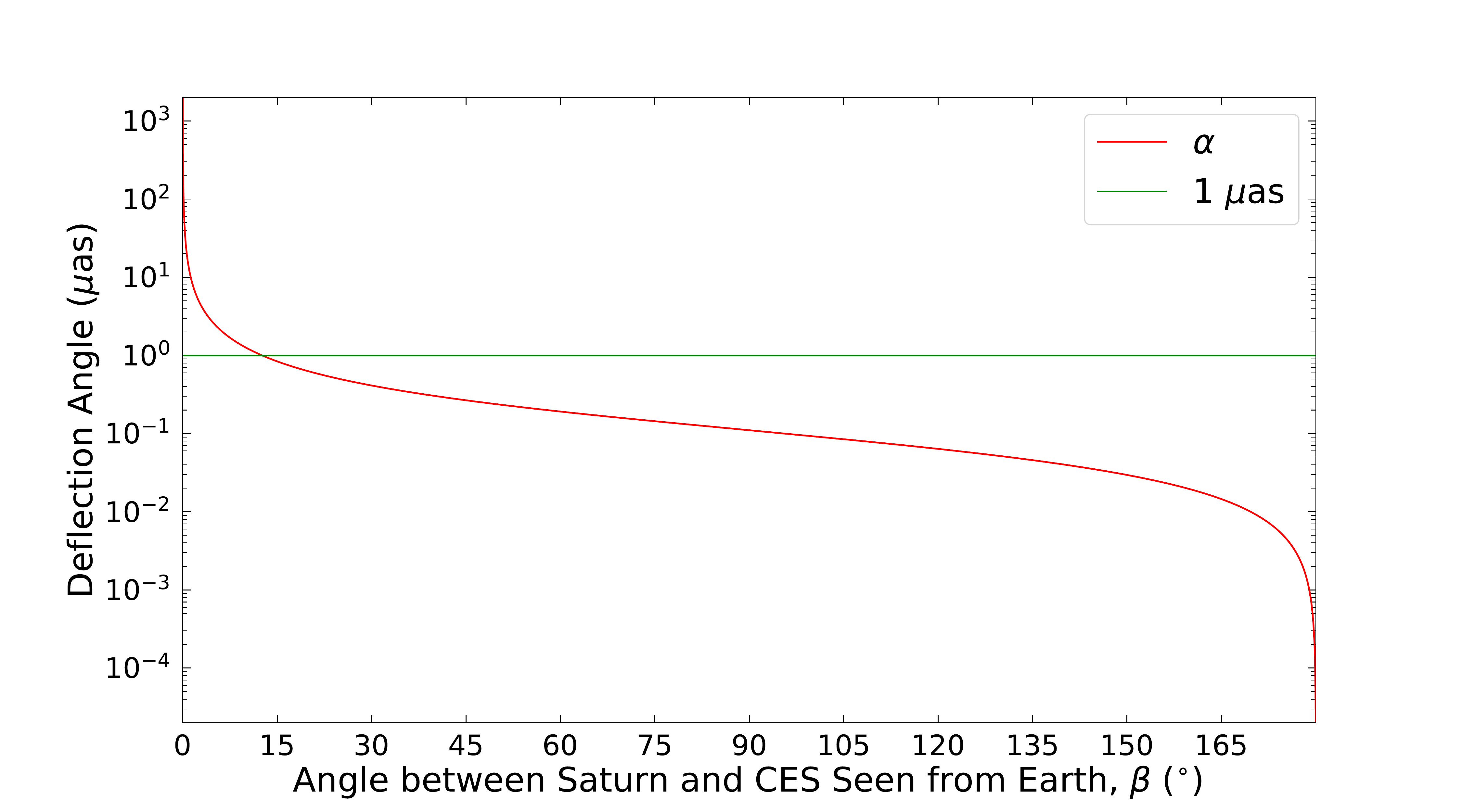}}
	\subfigure[Sun]{\includegraphics[width=0.49\textwidth]{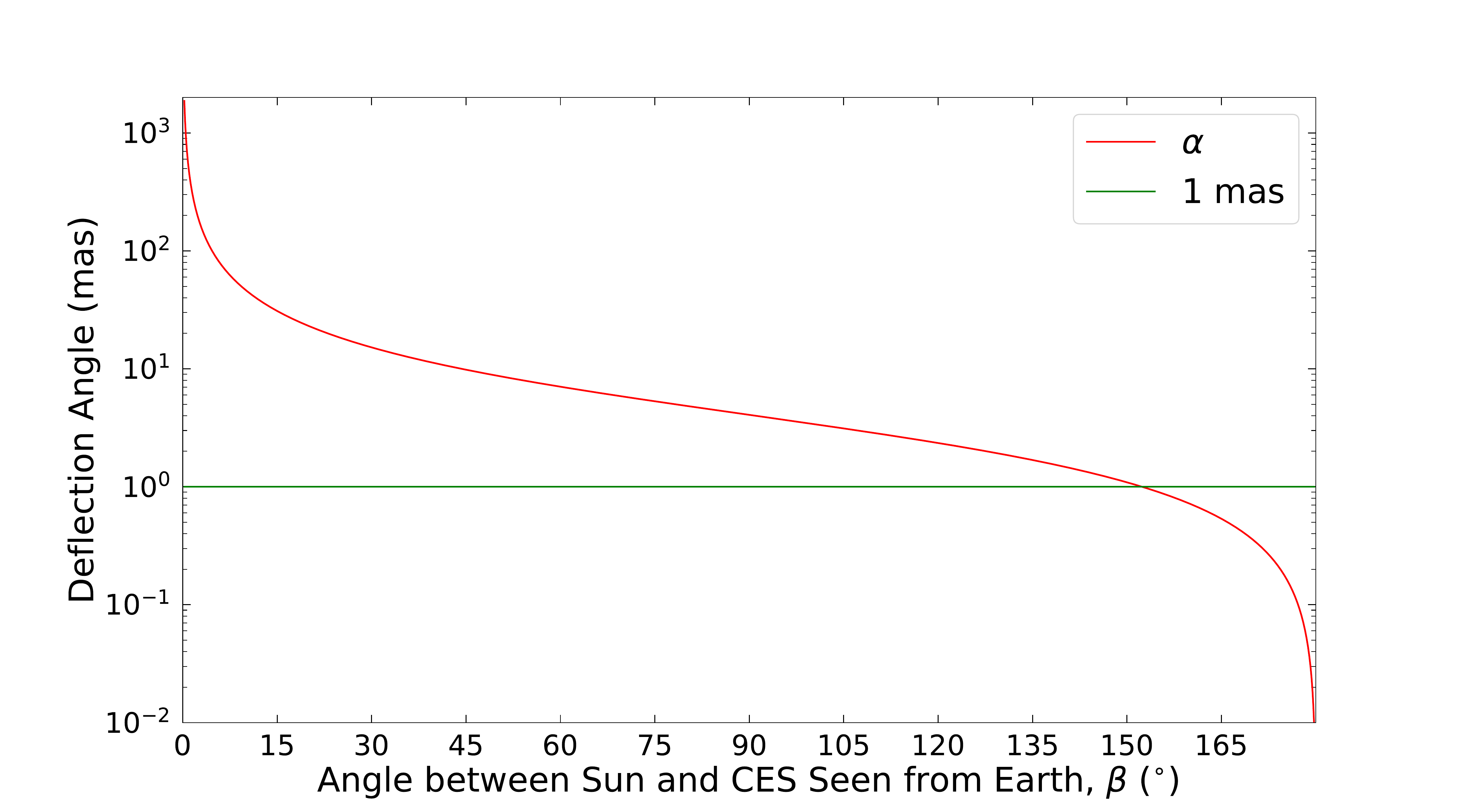}}	
	\caption{$\alpha$ as a function of $\beta$. The green lines denote a deflection angle of 1 $\mu$as for Jupiter and Saturn, and 1 mas for the Sun. The maximum and minimum distances roughly correspond to 5.2 (the semimajor axis of Jupiter's orbit) $\pm$ 1.0 au in panels (a) and (b), and 9.5 (the semi-major axis of Saturn's orbit) $\pm$ 1.0 au in panels (c) and (d), respectively.}
	\label{fig-alpha-position-rely}
\end{figure}

Light will bend when it passes through the gravitational field of a massive body, i.e. the massive body acts a gravitational lens (see the schematic diagram in Figure \ref{fig-skeptical}). The total deflection angle, $\alpha$, under the PPN formalism reads \citep{Cowling1984, Will1993, Ni2017}:
\begin{equation}\label{equ:alpha2}
	\alpha  = (1+
	\gamma)\frac{GM}{c^2 b}(\cos \theta_1 - \cos \theta_0),
\end{equation}
where $\theta_0 \approx 180^{\circ}$ if the observed target is a distant CES, $b$ is the impact parameter, $G$ the gravitational constant, $M$ the mass of the lens, and $c$ is the light speed. Because the angle between the $x$-axis and the vector $\overrightarrow{CE}$ is tiny, $b = r \sin \beta$ \citep[see][]{Cowling1984} and $\theta_1 \approx \alpha + \beta$, where $r$ is the distance from the Earth to the gravitational lens, and $\beta$ is the angle between each CES (in the absence of gravitational or aberrational bending) and the gravitational lens as seen by the observer. To better estimate $\alpha$ based on Equation (\ref{equ:alpha2}), several steps were taken as follows: (1) we replaced $\theta_1$ with the angle $\beta$, and calculated $\alpha$; and (2) we set $\theta_1 = \alpha + \beta$ and calculated $\alpha$. We found that $\alpha$ obtained from step (1) is equal to that derived from step (2) under a precision of 0.1 $\mu$as; therefore, only step (1) is conducted in this work. Figure \ref{fig-alpha-position-rely} shows examples of the deflection angle, $\alpha$, as a function of $\beta$. Jupiter can deflect light $\leq 85.5^{\circ}$ away from Jupiter by 1.0 $\mu$as, and $\leq 167.7^{\circ}$ away from Jupiter by 0.1 $\mu$as. Such large angles make Jupiter a nonnegligible light deflector in astrometry, like the Sun. Saturn can also deflect light by 0.1 $\mu$as if $\beta \leq 107.4^{\circ}$.

Here is a brief description of the correlator routine. The relative time delay between the antenna pairs (marked with the subscripts ``1" and ``2") caused by the Sun, the planets, the Moon, and the large satellites of Jupiter, Saturn, and Neptune was deducted in the process of the data correlator \citep[see][]{Pertit-Luzum2010}. It estimated the relative time delay under the Barycentric Celestial Reference System (BCRS) with $\gamma = 1$ as:
\begin{equation}\label{equ:time-delay}
	\Delta T_{\mathrm{grav},J} =(1+
	\gamma)\frac{GM_J}{c^3} \ln \frac{|\vec{R}_{1,J}| + \vec{K}\cdot\vec{R}_{1,J}}{|\vec{R}_{2,J}| + \vec{K}\cdot\vec{R}_{2,J}},
\end{equation}
where the subscript ``$J$" denotes the $J$-th gravitational body (e.g., Jupiter, Saturn, etc.), $\vec{R}_{i,J}$ is the vector from the $J$-th gravitational body to the $i$-th antenna, and $\vec{K}$ the unit vector from the barycenter to the source in the absence of gravitational or
aberrational bending. Both of these values are in the BCRS. The barycentric time delay in BCRS was then scaled to a terrestrial time interval and be deducted \citep[for more details, see][]{Pertit-Luzum2010}. The second-order PPN formalism effect also needs to be accounted for when light passes very close to the Sun \citep[for details, see][]{Richter-Matzner1983, Hellings1986, Pertit-Luzum2010}; however, as the nearest $b$ is $\sim$ 94 $R_{\odot}$ (in epoch 4), we did not include this effect in our calculations.

The data we received had been processed by the DiFX correlator, which had deducted the relative time delay caused by the Sun, the planets, the Moon, and the large satellites of Jupiter, Saturn, and Neptune. Therefore, the obtained CES positions using the AIPS software correspond to the positions after deducting the gravitational deflection effect with $\gamma = 1$. This methodology is equivalent to calculating the deflection angles caused by corresponding lenses using Equation (\ref{equ:alpha2}) and deducting them from the geometric optics approximation, where $\theta_0 \approx 180^{\circ}$. This correspondence can also been seen in \cite{Fomalont+2009}.

\subsubsection{Theoretical Results of the Deflection Angle}\label{sec:result-theory}

The planetary ephemerides in DE438 and the corresponding satellite ephemerides provided by JPL\footnote{For the ephemerides, see \url{https://naif.jpl.nasa.gov/pub/naif/generic_kernels/spk/} \citep{Folkner+2014}. The time series for each CES observation was extracted from the corresponding observational file in the form of `.sum' (which summarizes the start and stop times of each scan), with an accuracy of 1 second.} were used to calculate the position of the Sun, the planets and the satellites with PyEphem \citep[see][]{Rhodes2011}. The errors associated with the ephemerides are discussed in detail in Section \ref{sec:uncertaincy-emp}.

Table \ref{tab:alpha-theory} lists the theoretical values of the relative positions between the CES pairs in each epoch. In this work, Jupiter dominates the contribution to relative positions in R.A. The deflection by the Sun cannot be ignored. 

\begin{deluxetable}{ccccccccccccc}
	\tablecolumns{13}
	\tabletypesize{\small}
	\setlength\tabcolsep{2.8pt}
	\tablecaption{Theoretical Relative Positions of the Two CES Pairs during the Four Epochs
	\label{tab:alpha-theory}}
	%\begin{tabular}
	\tablehead{
		\colhead{Objects} & \colhead{Index} & \multicolumn{5}{c}{In the Direction of R.A.} & & \multicolumn{5}{c}{In the Direction of Decl.} \\
		\cline{3-7}  \cline{9-13}		
		\colhead{} & \colhead{} & \colhead{Epoch 1} & \colhead{Epoch 2} & \colhead{Epoch 3} & \colhead{Epoch 4} & \colhead{DR\tablenotemark{a}} & &  \colhead{Epoch 1} & \colhead{Epoch 2} & \colhead{Epoch 3} & \colhead{Epoch 4} & \colhead{DR} \\
		\colhead{} & \colhead{} & \colhead{($\mu$as)} & \colhead{($\mu$as)} & \colhead{($\mu$as)} & \colhead{($\mu$as)} & \colhead{($\mu$as)} & &  \colhead{($\mu$as)} & \colhead{($\mu$as)} & \colhead{($\mu$as)} & \colhead{($\mu$as)} & \colhead{($\mu$as)}
	}
	\startdata
    All\tablenotemark{b}	&	C1--C2	&	348.6 	&	97.5 	&	-492.6 	&	-106.8 	&	841.2 	&	&	258.5 	&	761.2 	&	317.3 	&	-900.0 	&	1661.3 	\\
    &	C1--C3	&	420.8 	&	169.4 	&	-420.1 	&	707.8 	&	1127.9 	&	&	251.6 	&	749.6 	&	300.0 	&	-921.3 	&	1670.9 	\\
    %         &	C2-C3	&	72.1 	&	71.9 	&	72.5 	&	814.6 	&	742.7 	&	&	-6.8 	&	-11.7 	&	-17.3 	&	-21.3 	&	14.4 	\\
    \hline
    Jupiter &	C1--C2	&	375.0 	&	123.8 	&	-466.0 	&	0.0 	&	841.0 	&	&	266.1 	&	771.0 	&	329.3 	&	-0.2 	&	771.2 	\\
    &	C1--C3	&	344.2 	&	90.4 	&	-501.1 	&	0.2 	&	845.3 	&	&	284.2 	&	784.9 	&	338.4 	&	-0.2 	&	785.1 	\\
    %         &	C2-C3	&	-30.8 	&	-33.4 	&	-35.1 	&	0.2 	&	35.3 	&	&	18.1 	&	13.9 	&	9.1 	&	0.0 	&	18.1 	\\
    \hline
    Sun &	C1--C2	&	-25.9 	&	-26.2 	&	-26.4 	&	-106.7 	&	80.8 	&	&	-7.1 	&	-9.3 	&	-11.6 	&	-899.7 	&	892.6 	\\
    &	C1--C3	&	76.7 	&	78.7 	&	80.8 	&	707.6 	&	630.9 	&	&	-31.6 	&	-34.6 	&	-37.7 	&	-921.0 	&	889.4 	\\
    %         &	C2-C3	&	102.6 	&	104.9 	&	107.2 	&	814.3 	&	711.7 	&	&	-24.4 	&	-25.2 	&	-26.1 	&	-21.3 	&	4.8 	\\
    \hline
    Others
    &	C1--C2	&	-0.5 	&	-0.2 	&	-0.2 	&	0.0 	&	0.5	&	&	-0.5 	&	-0.5 	&	-0.5 	&	-0.1 	&	0.4	\\
    &	C1--C3	&	-0.2 	&	0.2 	&	0.2 	&	0.1 	&	0.4	&	&	-1.0 	&	-0.7 	&	-0.7 	&	-0.1 	&	0.9	\\
    %         &	C2-C3	&	0.3 	&	0.4 	&	0.4 	&	0.1 	&	0.3	&	&	-0.5 	&	-0.3 	&	-0.3 	&	0.0 	&	0.5	\\ 
	\enddata
\tablenotetext{a}{The dynamic ranges among the four epochs.}
\tablenotetext{b}{Includes the deflection angle caused by the Sun, Mercury, Venus, Mars, Jupiter, Saturn, Uranus, and Neptune, and also the Moon and Ganymede (the most mossive satellite of Jupiter). Pluto is not included because the ephemerides data ``plu058'' only extend to 00:00:000 on 2015 December 1.}
\end{deluxetable}

In addition, other planets (excluding the Earth) and the two aforementioned satellites (the Moon and Ganymede) were also included in the theoretical values (see ``Others'' in Table \ref{tab:alpha-theory}). The values in the row ``All'' now include the contributions from ``Jupiter'', the ``Sun'',　and ``Others''. The contributions of ``Others'' (absolute values) to the relative positions are less than $\sim$ 1.0 $\mu$as, and the contributions to the dynamical ranges across the four epochs they are all less than $\sim$ 0.9 $\mu$as. The corresponding contributions in R.A. of interest for this work to both the relative positions and dynamical ranges are $\sim$ 0.5 $\mu$as, which are far lower than those of Jupiter and the Sun. Other satellites were ignored because even the contributions of Ganymede and the Moon are small.

Comparing the theoretical dynamical ranges of the relative positions (see the ``DR'' column in Table \ref{tab:alpha-theory}) throughout all of the epochs with those resulting from the observations (see the ``Std" column in Table \ref{tab:alpha-result}), and assuming that all of the errors of the observations are ascribed to gravitational deflection, the rough accuracy of $\gamma$ is $\sim 0.061$ for CES pair C1--C2 and $\sim 0.026$ for C1--C3.

\subsection{Analysis of $\gamma$}\label{sec:result-gamma}

To further analyze the value and error of $\gamma$, we should set up a new physical quantity, $V_\mathrm{new}$, whose value is the sum of the observed relative positions in R.A. and the theoretical ones with $\gamma = 1$. The reasons for setting up $V_\mathrm{new}$ are as follows: (1) in the correlation process, it assumed $\gamma = 1$ \citep{Pertit-Luzum2010} in all of the general relativity computations \citep[in the form of the PPN approximation; e.g.,][]{Will1993, Will2015, Ni2017}; and (2) the calculation of the ephemerides was not conducted in this work. We used the positions of the Sun, the planets, and the satellites as calculated by PyEphem \citep{Rhodes2011}, and the ephemerides as provided by JPL \citep{Folkner+2014}. In their numerical calculations of the ephemerides, they assumed $\gamma = 1$ when a gravity model was involved. However, the ephemerides should be calculated based on long-term monitoring, comprising both ground-based and space-based observations \citep{Folkner+2014, Pitjeva-Pitjev2018, Fienga+2020}. Therefore, the assumption of $\gamma = 1$ should hardly influence the results of the positions of celestial bodies in the solar system in this work. For further discussion about the ephemerides, see Section \ref{sec:uncertaincy-emp}. It is worth noting that the theoretical value with $\gamma = 1$ is from those labeled as ``All'' in Table \ref{tab:alpha-theory} (see Section \ref{sec:result-theory}), because the correlator model covered the Sun, the planets, the Moon, and the large satellites of Jupiter, Saturn, and Neptune \citep{Pertit-Luzum2010}.

Specifically, $V_\mathrm{new}$ reads:
\begin{gather}\label{equ:alpha-multi-gamma}
V_\mathrm{new}(i) = V_\mathrm{obs}(i) + C_{C1,C3} + V_{\gamma=1}(i),\;\; \text{for}\;\; i = 1, 2, 3, 4;\\
V_\mathrm{new}(i) = V_\mathrm{obs}(i) + C_{C1,C2}  + V_{\gamma=1}(i),\;\; \text{for}\;\; i = 5, 6, 7, 8.
\end{gather}
The indexes of $i = 1, 2, 3, 4$ correspond to the four epochs of CES pair C1--C3, and $i = 5, 6, 7, 8$ correspond to those of C1--C2. $V_\mathrm{obs}$ denotes the observed relative positions (see Table \ref{tab:alpha-result}). $C_{C1,C2}$ and $C_{C1,C3}$, which remain constant across the four epochs, should also be considered here. $C_{C1,C3} = 0$ for $i = 5, 6, 7, 8$ and $C_{C1,C2} = 0$ for $i = 1, 2, 3, 4$. $V_{\gamma=1}$ represents the theoretical relative positions assuming $\gamma = 1$ (see the ``All'' column in Table \ref{tab:alpha-theory}).

\begin{figure}[!htb]
	\centering
	\includegraphics[width=0.9\textwidth]{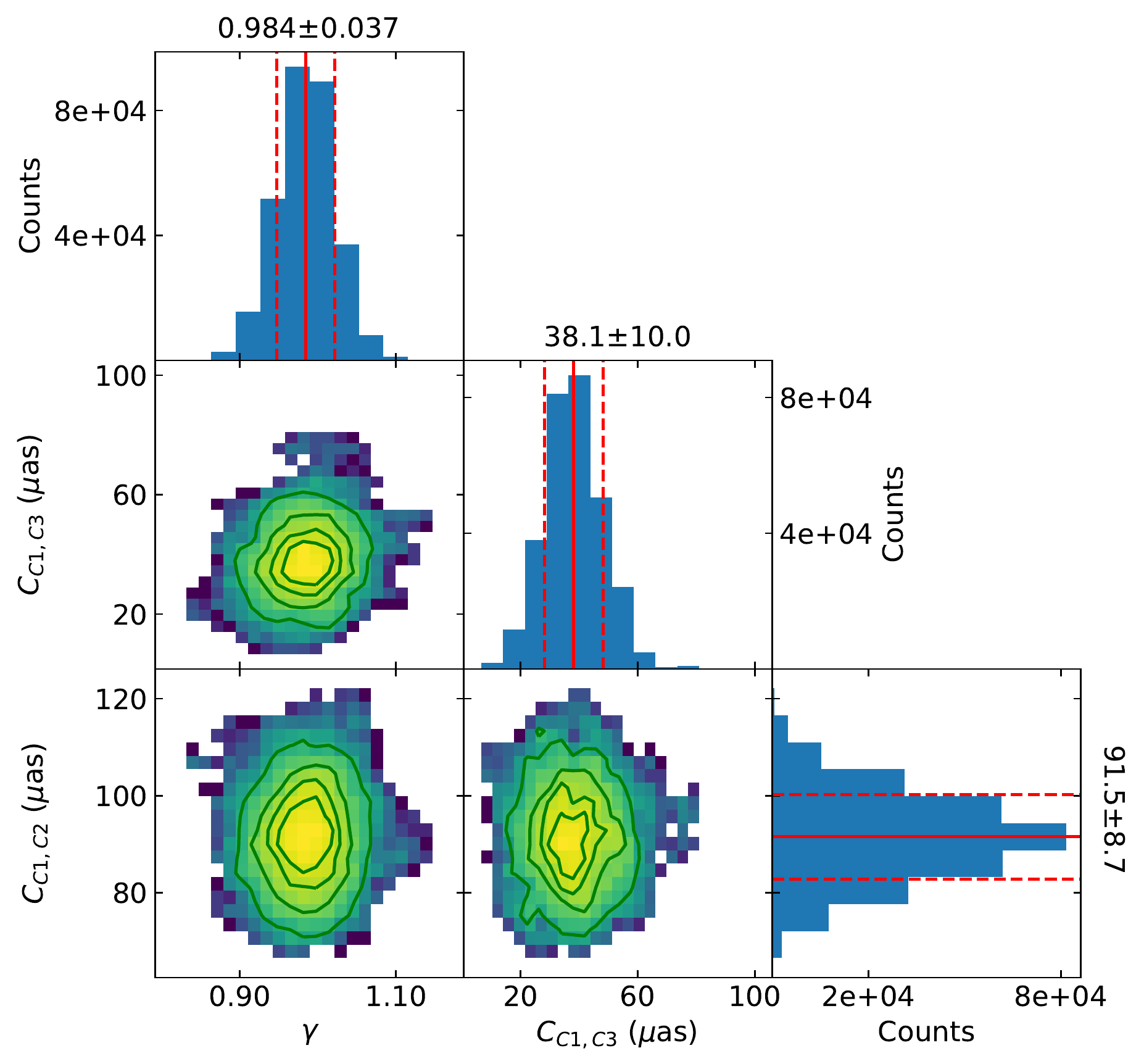}	
	\caption{Estimations of $\gamma$, $C_{C1,C2}$, and $C_{C1,C3}$, in units of $\mu$as using the MCMC method. The contours are 0.1, 0.3, 0.5, and 0.7 times of the peak values.}
	\label{fig-gamma-static}
\end{figure}

The theoretical relative position with a value of $\gamma$ is $V_{\gamma}$ (similar to $V_{\gamma = 1}$, but with different values of $\gamma$). Markov Chain Monte Carlo (MCMC) trials \citep{Metropolis+1953,Goodman-Weare2010} were used to estimate the maximum likely values of $\gamma$, $C_{C1,C2}$, and $C_{C1,C3}$ with the evaluation function, $\ln (\mathcal{L})$, as:
\begin{gather}\label{equ:alpha-multi-gamma}
\ln (\mathcal{L}) = -\frac{1}{2}\sum_{i=1}^{8}\frac{\left[V_\mathrm{new}(i) - V_{\gamma}(i)\right]^2}{\left[\sigma_\mathrm{obs}(i)\right]^2}.
\end{gather}
Similar to $V_{\mathrm{obs}}$, the formal errors of the observed relative positions, $\sigma_\mathrm{obs}(i)$, are listed in Table \ref{tab:alpha-result}. Consequently, we calculated 10 MCMC sampling chains, where the first 40\% were discarded. Finally, the total number of samples was 300\,065. Figure \ref{fig-gamma-static} shows the MCMC result for $\gamma = 0.984 \pm 0.037$, which is consistent with the value of $1.01 \pm 0.03$ in \citet{Fomalont-Kopeikin2003, Fomalont-Kopeikin2008}. For the results from \citet{Fomalont-Kopeikin2003, Fomalont-Kopeikin2008}, the formal accuracy of the measurement is also $\sim$ 20 $\mu$as and the smallest $b$ (with the lens being Jupiter) is $\sim$ 13 $R_J$ (slightly smaller than that in this work, i.e., $\sim$ 19 $R_J$).

\section{Discussion of Uncertainties from Other Factors}\label{sec:uncertaincy}

\subsection{The Impact of Plasma}\label{sec:uncertaincy-pla}

The deflection angle caused by the plasma in the Sun's corona could be larger than that caused by the solar gravitational field at 1 GHz when light grazes the solar surface \citep{Shapiro1964}. The electron density, $N_e$, in the solar corona or Jupiter's atmosphere is assumed to have the form: $N_e(r) = N_0 (R/r)^{A+2}$, where $r$ is the distance from the Sun or Jupiter and $R$ is their radius in units of m. $A$ is a number and $N_0$ is a constant with units of m$^{-3}$. The maximum deflection angle caused by plasma, $\alpha_p$, in units of $\mu$as reads as \citep[see][]{Kopeikin-Fomalont2002, Kopeikin-Makarov2007}:
\begin{equation}
	\alpha_p = 8.3 \times 10^{-6} (A+1) \frac{N_I}{b} \left(\frac{1}{\nu}\right)^2,
\end{equation}
where $\nu$ is in units of GHz and $b$ is in units of m, and the electron integral column density, $N_I$, along the line of sight is:
\begin{equation}
	N_I = \int^{r_0}_b \frac{N_e(r)rdr}{\sqrt{r^2-b^2}}+ \int^{r_1}_b \frac{N_e(r)rdr}{\sqrt{r^2-b^2}},
\end{equation}
where $b$ is the impact parameter, and $r_0$ and $r_1$---both are much greater than $b$---are the distances from the Sun or Jupiter to the CES and the Earth, respectively.

\begin{figure}[!htb]
	\centering
	\includegraphics[width=0.7\textwidth]{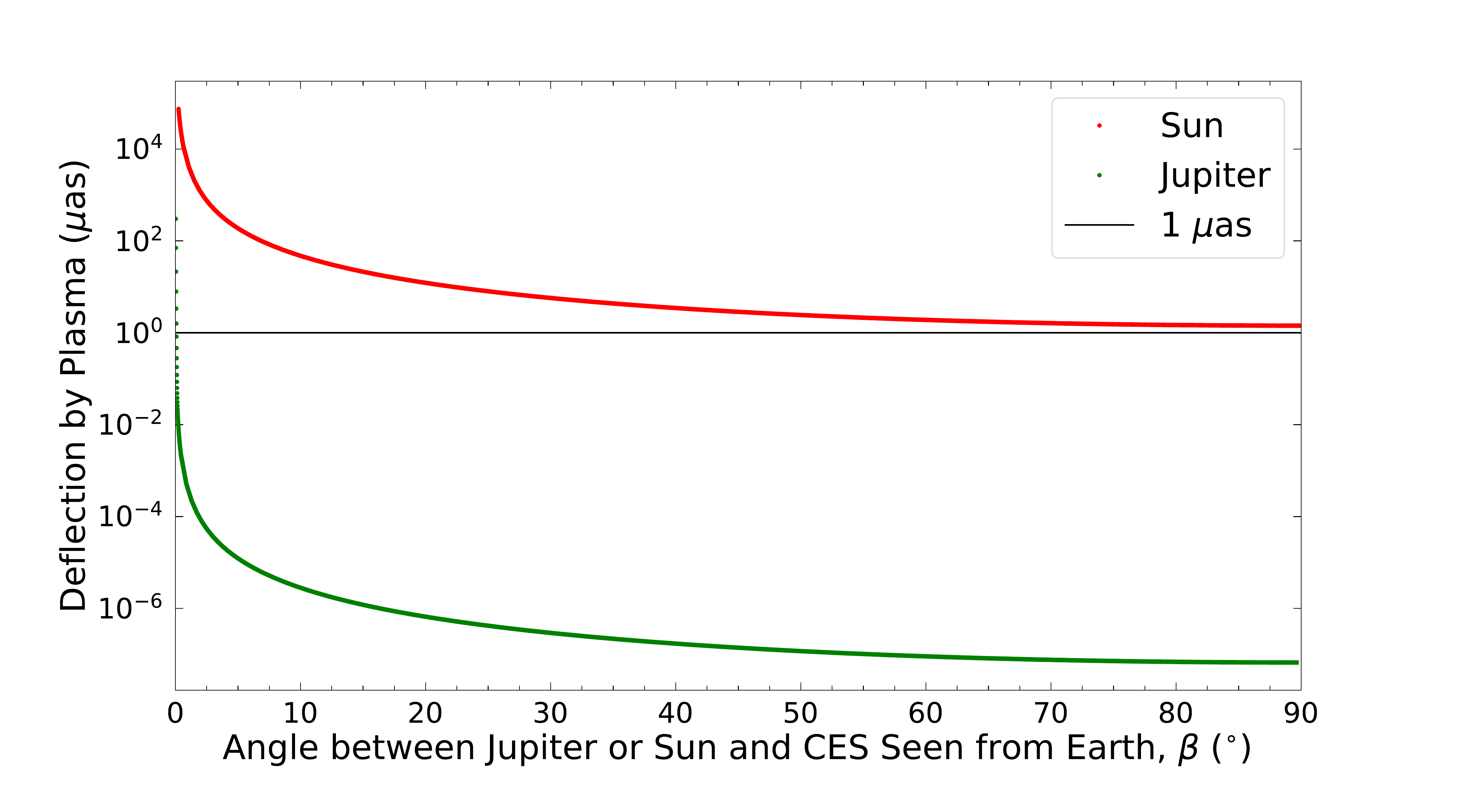}
	\caption{Deflection of light caused by plasma in the solar corona and Jupiter's atmosphere at $\nu = 15$ GHz. The distance from the Earth to Jupiter was set to a mean distance of $\sim$ 5.2 au.}
	\label{fig-alpha-plasma}
\end{figure}

The plasma of the solar corona has been well studied, but knowledge of plasma in Jupiter's atmosphere is scant. For the Sun, $N_0 = 0.57 \times 10^{12}$ m$^{-3}$ with $A = 0$, which was measured during 2011 and 2012, and this value can be considered a worst case scenario \citep{Soja+2014, Titov+2018}. To describe the electron content within 4 $R_{\odot}$ (corresponding to $\beta \sim 1.1^{\circ}$) from the solar center, additional terms are required \citep{Verma+2013,Soja+2014}. In this work, however, the targeted CESs are far from the solar center (the nearest $b$ is $\sim$ 94 $R_{\odot}$), thus the additional terms can be ignored. The deflection caused by the solar corona is plotted in Figure \ref{fig-alpha-plasma}. We note that the plotted values are overestimated by at least a factor of two when $\beta$ approaches $90^{\circ}$.

For Jupiter's atmosphere, it is hard to obtain comparatively accurate information about its plasma content from the few available observations. In this work, we have adopted the results reported by \cite{Ansher2001}, who used data from the plasma wave instruments on board the Galileo spacecraft. In that work, Jupiter's atmosphere was divided into two parts and was bounded by $\sim$ 20--30 $R_J$, where $R_J$ is the mean radius of Jupiter. In the inner region, $5 < r < 30$ $R_J$, $N_0 = 2.74 \times 10^{8}$ cm$^{-3}$ and $A = 4.55 \pm 0.05$, while in the outer region, $20 < r < 140$ $R_J$, $N_0 = 250$ cm$^{-3}$ and $A = 0.14 \pm 0.01$. The change in density values may deviate from power-law profiles by at least a factor of two. To estimate the maximum bending of light caused by plasma, we extended the two profiles to $r > 1$ $R_J$ and summed $\alpha_p$ from them. The result is mapped in Figure \ref{fig-alpha-plasma}.

Figure \ref{fig-alpha-plasma} indicates that the impact of plasma (i.e. $\alpha_p$) in Jupiter's atmosphere is far less than that in the solar corona. Both of them decrease much more steeply than the deflection angle caused by the gravitational field as a function of $\beta$ (see Figure \ref{fig-alpha-position-rely}). The deflection by Jupiter's atmosphere would be less than 1 $\mu$as as long as $\beta > 5'$. Therefore, choosing Jupiter as a gravitational lens can greatly reduce the limitation of the measured accuracy of the gravitational deflection angle posed by the solar corona.

In this work, the closest distance of light from the three CESs to Jupiter occurred in epochs 1--3, with an impact parameter $b \sim 19$ $R_J$ (see Figure \ref{fig-design}). The corresponding value of $\alpha_p$ is $\sim$ 0.3 $\mu$as, less than 1 $\mu$as. For the Sun, in epochs 1--3, the impact parameter was $b > 209$ $R_{\odot}$ for the three CESs, and the corresponding $\alpha_p$ was $\sim 1.5$ $\mu$as. In epoch 4, $94 < b < 95$ $R_{\odot}$ for the three CESs, with the corresponding $\alpha_p$ being about 7.3--8.0 $\mu$as. Moreover, these values are overestimated, and the relative bending angles of the CES pairs are likely much smaller. In addition, they are both less than the formal accuracy by a factor of at least $\sim$ 4. Therefore, they can be ignored. 

Because plasma in Jupiter's atmosphere near its surface is a controversial issue, a more detailed discussion of the electron content in the inner region is presented here. The data in the inner region from the work of \citet{Ansher2001} are from a different study without citation. Additionally, $N_0 = 2.74 \times 10^{14}$ m$^{-3}$, i.e., $N_e(r) = 2.74 \times 10^{8} (R/r)^{6.55}$ cm$^{-3}$, which is much larger than that at the solar surface. The results from the first Galileo mission and Voyager 2's Radio Occultation Experiment showed that $N_e$ close to Jupiter's surface (i.e., several hundred to a thousand kilometers above Jupiter's surface) is $\sim 1 \times 10^{11}$ m$^{-3}$ \citep{Hinson+1997,Hinson+1998}. If we assume that $N_e \sim 1 \times 10^{11}$ m$^{-3}$ at $r = 1.0$ $R_J$, and $N_e = 2.5 \times 10^8 (R/r)^{2.14}$ for $r > 20$ $R_J$ \citep{Ansher2001}, and that profile of $N_e$ in the range $1 < r < 20$ $R_J$ obeys a power law distribution, $N_e(r) = 1 \times 10^{11} (R/r)^{4.1}$. The corresponding $\alpha_p$ value for the range $1 < b < 10$ $R_J$ would be 1--2 orders of magnitude lower than the value plotted in Figure \ref{fig-alpha-plasma}. Therefore, the value of $\alpha_p$ plotted in Figure \ref{fig-alpha-plasma} for Jupiter may be significantly overestimated, especially when $b$ is small.

\subsection{The Impact of the Uncertainties of the Positions of the Planets and the Sun}\label{sec:uncertaincy-emp}

The orbital accuracies of tens of kilometers in JPL DE430 or DE431, released in 2014 for Jupiter and Saturn \citep{Folkner+2014}, correspond to positional accuracies of from about several mas to a dozen mas. From DE430 to DE440, released in 2021, the positional accuracies have improved to below 1 mas \citep{Park+2021}. The ephemerides adopted in this work are from DE438, released in 2018, which have been tuned for the flight project's target body, e.g., the Juno mission \citep{Bolton+2017, Park+2021}. Therefore, the positional accuracies of the planets in DE438 are better than those in DE430 or DE431. Therefore, the accuracy of $\beta$ was set to 10 mas for the cases of both Jupiter and Saturn. 

\begin{figure}[!htb]
	\centering
	\subfigbottomskip=-0.2cm
	\subfigcapskip=-0.2cm
	\subfigure[Jupiter with Maximum Distance from Earth]{\includegraphics[width=0.49\textwidth]{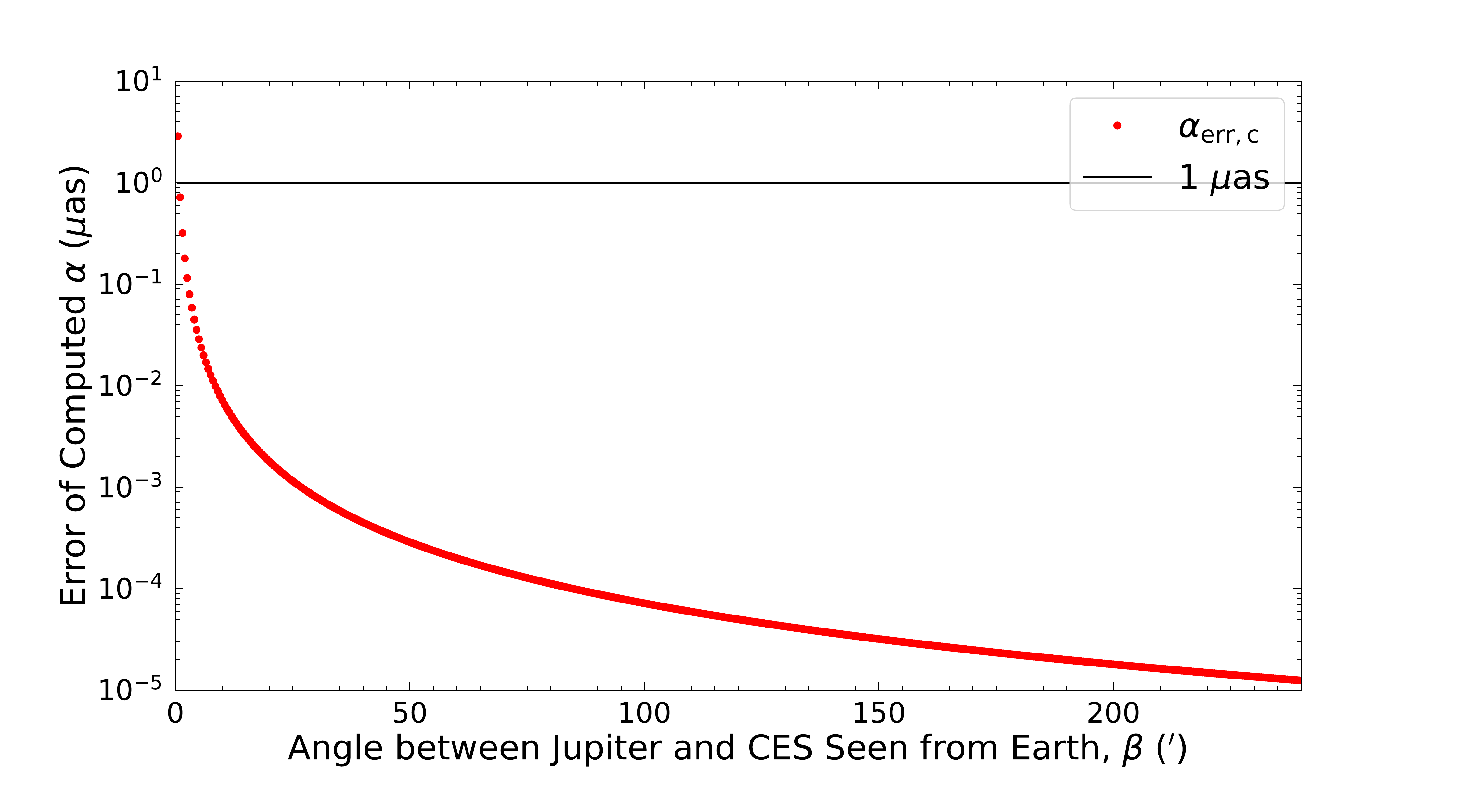}}
	\subfigure[Jupiter with Minimum Distance from Earth]{\includegraphics[width=0.49\textwidth]{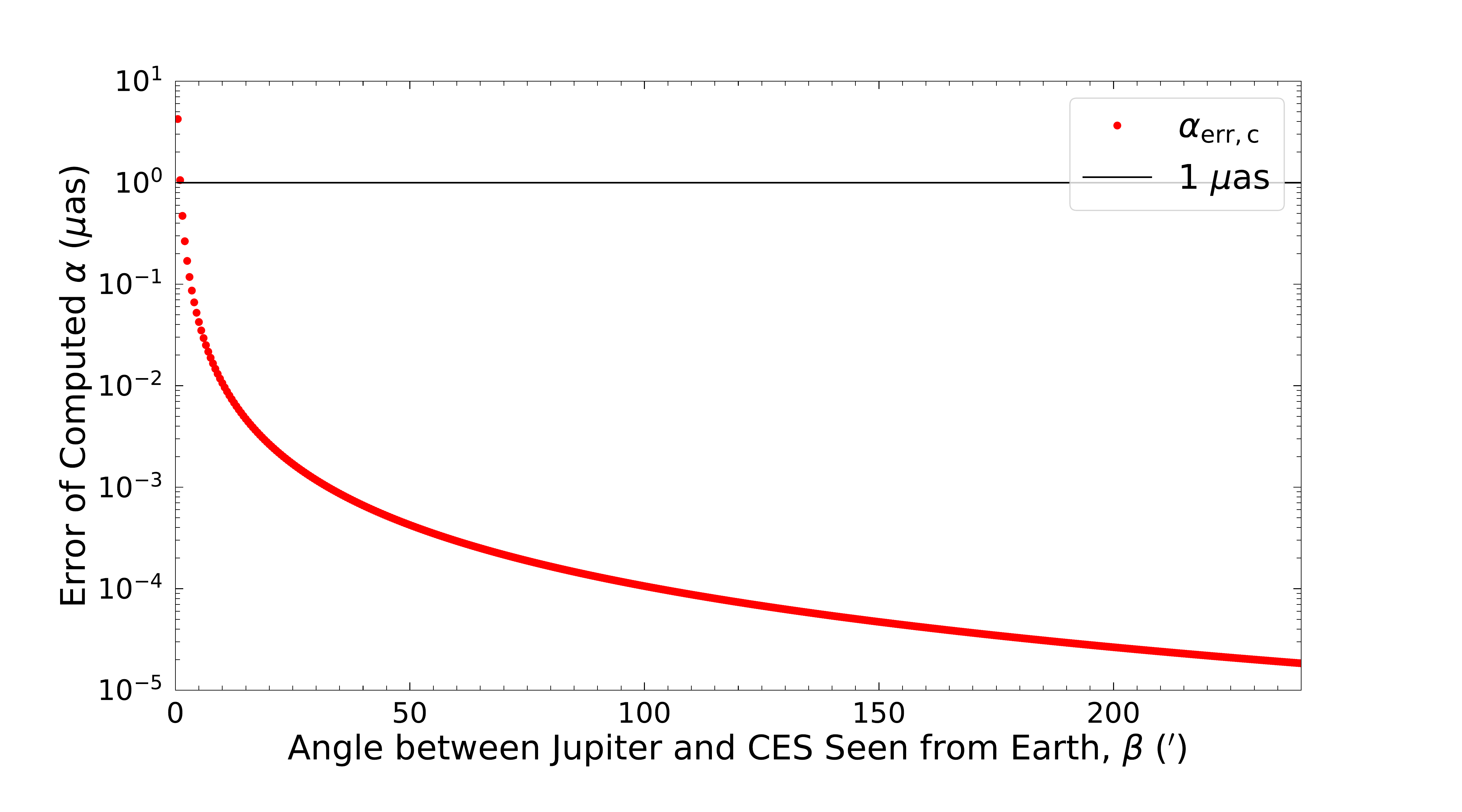}}	
	\subfigure[Saturn with Maximum Distance from Earth]{\includegraphics[width=0.49\textwidth]{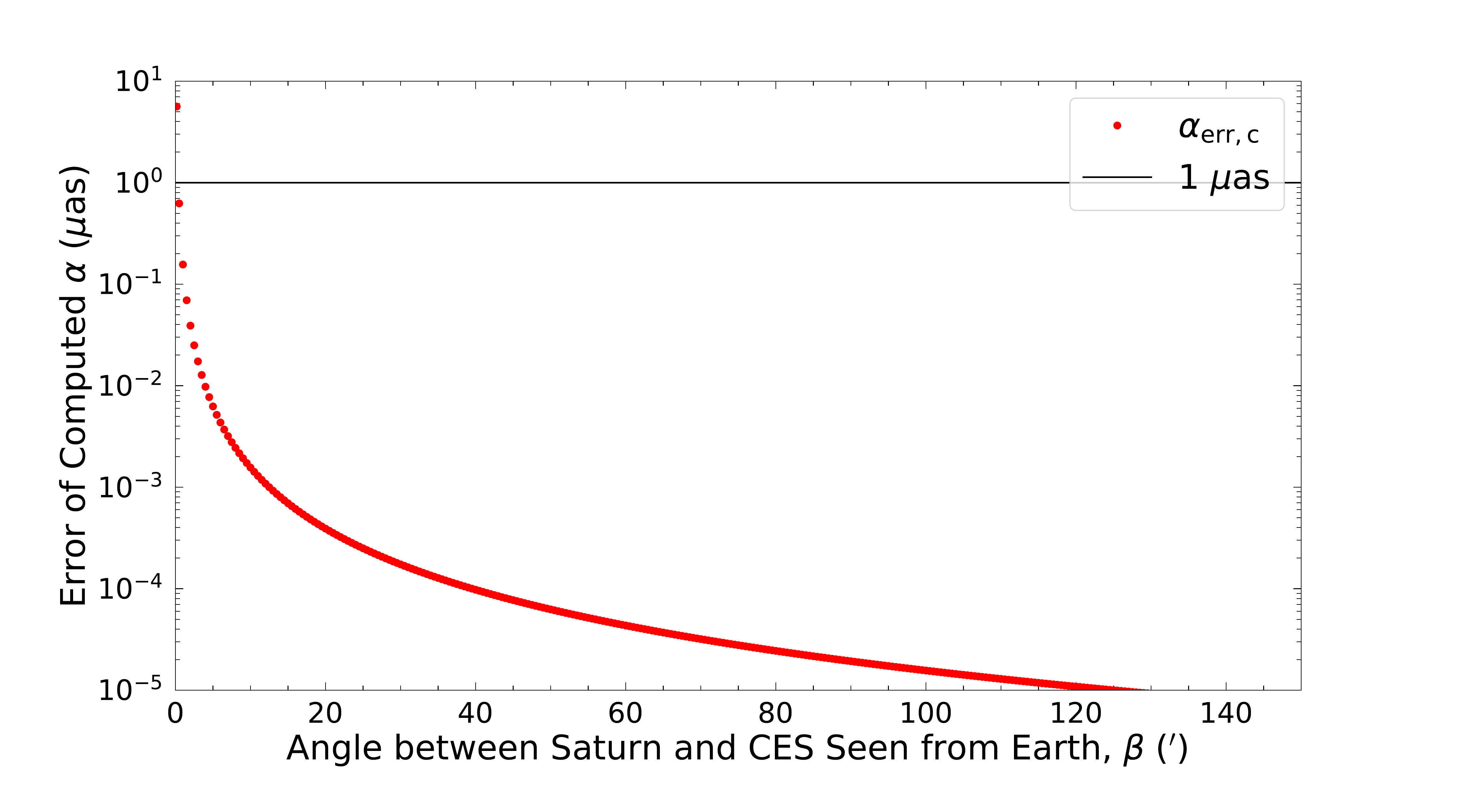}}
	\subfigure[Saturn with Minimum Distance from Earth]{\includegraphics[width=0.49\textwidth]{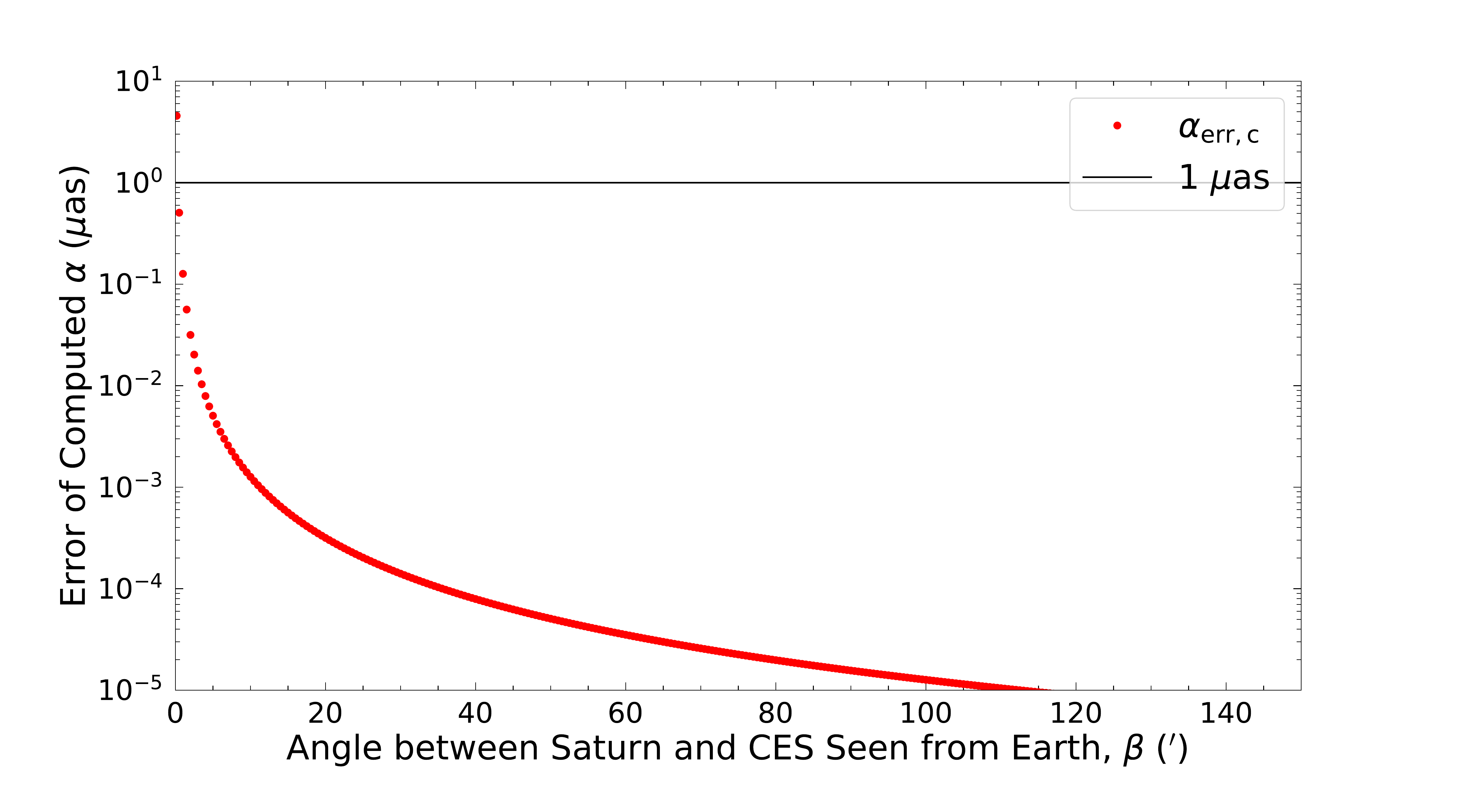}}
	\subfigure[Sun]{\includegraphics[width=0.49\textwidth]{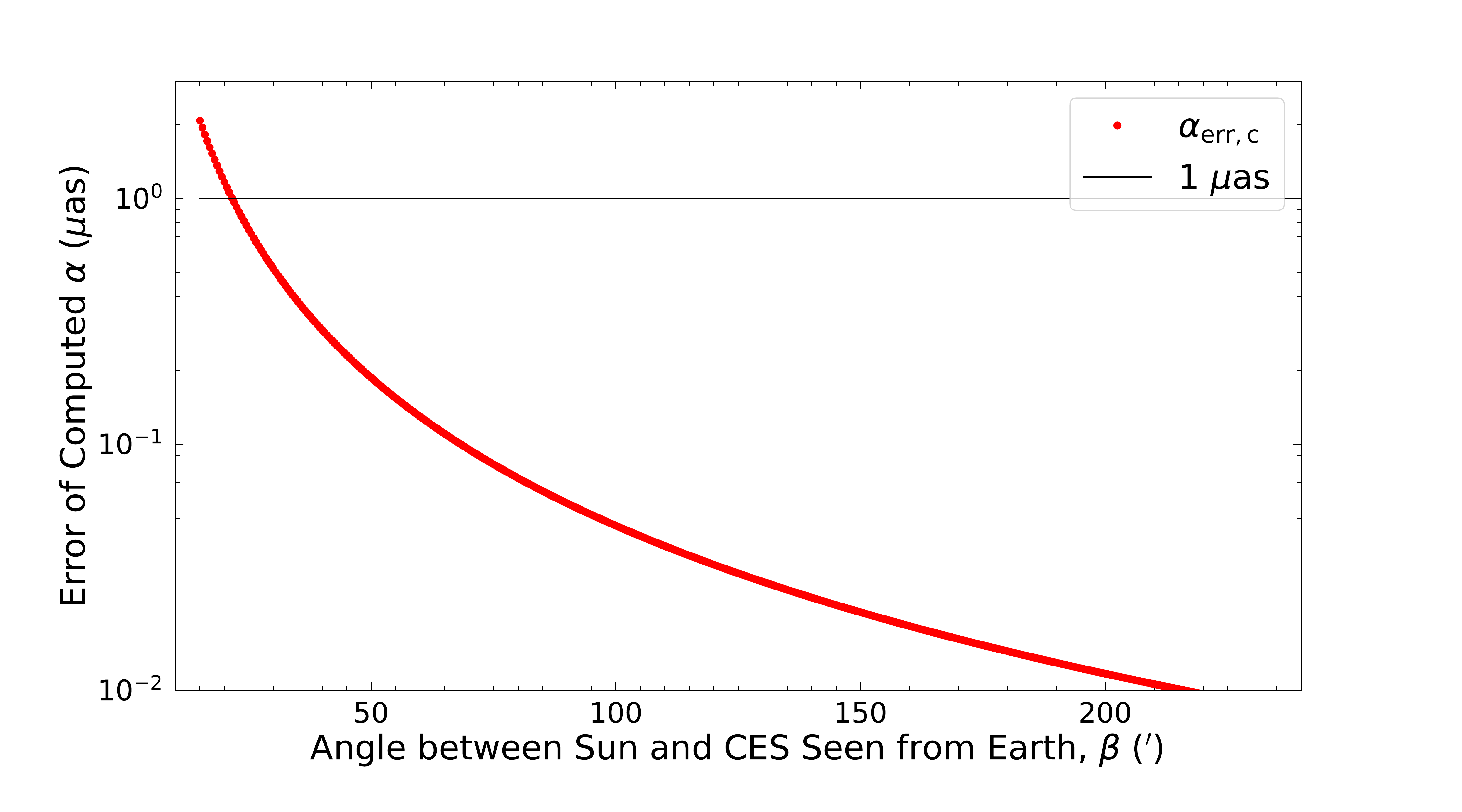}}	
	\caption{The contribution from the uncertainty of $\beta$ (see the text) to the deflection angle, $\alpha_{err,c}$.}
	\label{fig-alpha_compute_err}
\end{figure}

The orbit accuracies of the inner planets are $\sim$ 0.2 mas \citep{Folkner+2014, Folkner-Border2015}, and Earth's orbit is around the Sun to the first order \citep{Park+2021}. Here, the accuracy of $\beta$ was set to 0.5 mas for the cases of the Sun and the terrestrial planets. The contribution from the uncertainty of $\beta$ with the value set above was roughly calculated and plotted in Figure \ref{fig-alpha_compute_err}. 

It can be seen that the contribution from the uncertainty of $\beta$ to the deflection angle, $\alpha_{err,c}$, is small. For Jupiter, when it is farthest from the Earth, the contribution is less than 1 $\mu$as as long as the light is $\gtrsim 1.0'$ (corresponding to $b \gtrsim 3$ $R_J$) from Jupiter. The corresponding $\beta$ is $\gtrsim 1.5'$ (corresponding to $b \gtrsim 4$ $R_J$) when Jupiter is nearest to the Earth. For Saturn, the contribution is less than 1 $\mu$as as long as the light is $\gtrsim 1.0'$ (corresponding to $b \gtrsim 7$ $R_S$, where $R_S$ is the mean radius of Saturn) away from Saturn, no matter where it is. The corresponding $\beta$ is $\gtrsim 22.0'$ (corresponding to $b \gtrsim 1$ $R_{\odot}$) for the Sun. Therefore, in this work, the impact of the uncertainties of the positions of gravitational lenses can be ignored.

\section{Summary and Conclusion}\label{sec:summary}

We conducted four epochs of observations, on 2020 October 23--25 and 2021 February 5 (GST), of the relative positions between CESs J1925-2219 and two calibrators (J1928-2035 and J1923-2104) in order to see how they are impacted by Jupiter as well as other objects in the solar system. The dynamical ranges of the theoretical relative positions of CES pairs C1--C2 and C1--C3 in R.A. are 841.2 and 1127.9 $\mu$as, respectively, where the gravitational lenses include the Sun, Mercury, Venus, Mars, Jupiter, Saturn, Uranus, and Neptune, and also the Moon and Ganymede. Jupiter's impact dominates the contributions to the dynamical ranges. We directly mapped the CESs to obtain their relative positions. Due to the very low elevation angles, the error in Decl. is poor. The formal accuracy in R.A. is about 20 $\mu$as. The measured standard deviations of the relative positions across the four epochs are 51.0 and 29.7 $\mu$as for C1--C2 and C1--C3 in R.A., respectively. The rough accuracy of $\gamma$ is therefore $\sim 0.061$ for CES pair C1--C2 and $\sim 0.026$ for C1--C3. The measured value of $\gamma$, determined from both of the CES pairs, is $0.984 \pm 0.037$, as determined from the MCMC trials. This result is comparable to the latest published results that consider Jupiter as a gravitational lens, i.e., $1.01 \pm 0.03$ \citep[see][]{Fomalont-Kopeikin2003, Fomalont-Kopeikin2008}. The influence of plasma and the uncertainties of positions of gravitational lenses caused by the uncertainties of the ephemerides can be ignored.

\acknowledgments

We would like to thank the referee for reviewing the paper carefully and the constructive comments that improves this manuscript. This work was sponsored by the Natural Science Foundation of Jiangsu Province (Grants No. BK20210999), the NSFC Grants Nos. 11933011 and 11873019, the Key Laboratory for Radio Astronomy, CAS. 

\facility{VLBA}
\software{DiFX \citep{Deller+2007}, AIPS \citep{van-Moorsel+1996}, PyEphem \citep{Rhodes2011}, Astropy \citep{Astropy2013,Astropy2018}, Matplotlib \citep{Matplotlib2007}, Numpy \citep{Numpy2020}, Pandas \citep{Pandas2021}, GeoGebra (\url{https://www.geogebra.org}, Copyright \copyright~International GeoGebra Institute, 2013).}

\setcounter{table}{0}
\renewcommand{\thetable}{A\arabic{table}}

\bibliographystyle{plainnat}
\bibliography{liyj_v1_bib}
	
\end{CJK*}
\end{document}